\documentclass[acmsmall]{acmart}

\usepackage{algpseudocode}
\usepackage{amsmath}  
\usepackage[ruled,linesnumbered]{algorithm2e}

\usepackage{multirow}
\usepackage{algpseudocode}
\usepackage{tikz}
\newcommand{\circnumber}[1]{\lower.75ex\hbox{\tikz\draw (0pt, 0pt)%
    circle (.47em) node {\makebox[.15em][c]{\small #1}};}}
\newcommand{\ballnumber}[1]{\lower.75ex\hbox{\tikz\fill(0pt, 0pt)%
    circle (.5em) node {\makebox[.15em][c]{\small \textcolor{white}{#1}}};}}
\newcommand{\dcircnumber}[1]{\lower.75ex\hbox{\tikz\draw(0pt, 0pt)%
    circle (.47em) circle (.37em) node {\makebox[.15em][c]{\small #1}};}}

\newcommand{\emptycirc}[1]{\lower.75ex\hbox{\tikz\draw (0pt, 0pt)%
    circle (.47em) node {\makebox[.15em][c]{\small \textcolor{white}{#1}}};}}
\newcommand{\emptyball}[1]{\lower.75ex\hbox{\tikz\fill(0pt, 0pt)%
    circle (.5em) node {\makebox[.15em][c]{\small #1}};}}
\newcommand{\emptydcirc}[1]{\lower.75ex\hbox{\tikz\draw(0pt, 0pt)%
    circle (.47em) circle (.37em) node {\makebox[.15em][c]{\small \textcolor{white}{#1}}};}}
            
\usepackage{listings, xcolor}
\definecolor{verylightgray}{rgb}{.97,.97,.97}
\lstdefinelanguage{Solidity}{
  keywords=[1]{anonymous, assembly, assert, balance, break, call, callcode, case, catch, class, constant, continue, constructor, contract, debugger, default, delegatecall, delete, do, else, emit, event, experimental, export, external, false, finally, for, function, gas, if, implements, import, in, indexed, instanceof, interface, internal, is, length, library, log0, log1, log2, log3, log4, memory, modifier, new, payable, pragma, private, protected, public, pure, push, require, return, returns, revert, selfdestruct, send, solidity, storage, struct, suicide, super, switch, then, this, throw, transfer, true, try, typeof, using, value, view, while, with, addmod, ecrecover, keccak256, mulmod, ripemd160, sha256, sha3}, 
  keywordstyle=[1]\color{blue}\bfseries,
  keywords=[2]{address, bool, byte, bytes, bytes1, bytes2, bytes3, bytes4, bytes5, bytes6, bytes7, bytes8, bytes9, bytes10, bytes11, bytes12, bytes13, bytes14, bytes15, bytes16, bytes17, bytes18, bytes19, bytes20, bytes21, bytes22, bytes23, bytes24, bytes25, bytes26, bytes27, bytes28, bytes29, bytes30, bytes31, bytes32, enum, int, int8, int16, int24, int32, int40, int48, int56, int64, int72, int80, int88, int96, int104, int112, int120, int128, int136, int144, int152, int160, int168, int176, int184, int192, int200, int208, int216, int224, int232, int240, int248, int256, mapping, string, uint, uint8, uint16, uint24, uint32, uint40, uint48, uint56, uint64, uint72, uint80, uint88, uint96, uint104, uint112, uint120, uint128, uint136, uint144, uint152, uint160, uint168, uint176, uint184, uint192, uint200, uint208, uint216, uint224, uint232, uint240, uint248, uint256, var, void, ether, finney, szabo, wei, days, hours, minutes, seconds, weeks, years},  
  keywordstyle=[2]\color{teal}\bfseries,
  keywords=[3]{block, blockhash, coinbase, difficulty, gaslimit, number, timestamp, msg, data, gas, sender, sig, value, now, tx, gasprice, origin},  
  keywordstyle=[3]\color{violet}\bfseries,
  identifierstyle=\color{black},
  sensitive=false,
  comment=[l]{//},
  morecomment=[s]{/*}{*/},
  commentstyle=\color{gray}\ttfamily,
  stringstyle=\color{red}\ttfamily,
  morestring=[b]',
  morestring=[b]"
}
\usepackage{soul} 
\usepackage{color, xcolor} 

\AtBeginDocument{%
  }


\setcopyright{cc}
\setcctype{by}
\acmDOI{10.1145/3728921}
\acmYear{2025}
\acmJournal{PACMSE}
\acmVolume{2}
\acmNumber{ISSTA}
\acmArticle{ISSTA046}
\acmMonth{7}
\received{2025-02-24}
\received[accepted]{2025-03-31}

\begin{document}

\title{Copy-and-Paste? Identifying EVM-Inequivalent Code Smells in Multi-chain Reuse Contracts}


\author{Zexu Wang}
\orcid{0009-0004-1439-2989}
\affiliation{%
  \institution{Sun Yat-sen University}
  \city{Zhuhai}
  \country{China}
}
\affiliation{%
  \institution{Peng Cheng Laboratory}
  \city{Shenzhen}
  \country{China}
}
\email{wangzx97@mail2.sysu.edu.cn}

\author{Jiachi Chen}
\orcid{0000-0002-0192-9992}
\affiliation{%
  \institution{Sun Yat-sen University}
  \city{Zhuhai}
  \country{China}
}
\affiliation{%
  \institution{Guangdong Engineering Technology Research Center of Blockchain}
  \city{Zhuhai}
  \country{China}
}
\email{chenjch86@mail.sysu.edu.cn}

\author{Tao Zhang}
\orcid{0000-0002-6272-4069}
\affiliation{%
  \institution{Macau University of Science and Technology}
  \city{Macau}
  \country{China}
}
\email{tazhang@must.edu.mo}

\author{Yu Zhang}
\orcid{0000-0003-2040-5059}
\affiliation{%
  \institution{Harbin Institute of Technology}
  \city{Harbin}
  \country{China}
}
\affiliation{%
  \institution{Peng Cheng Laboratory}
  \city{Shenzhen}
  \country{China}
}
\email{yuzhang@hit.edu.cn}

\author{Weizhe Zhang}
\orcid{0000-0003-4783-876X}
\affiliation{%
  \institution{Harbin Institute of Technology}
  \city{Harbin}
  \country{China}
}
\affiliation{%
  \institution{Peng Cheng Laboratory}
  \city{Shenzhen}
  \country{China}
}
\email{wzzhang@hit.edu.cn}

\author{Yuming Feng}
\orcid{0000-0001-8922-0496}
\affiliation{%
  \institution{Peng Cheng Laboratory}
  \city{Shenzhen}
  \country{China}
}
\email{fengym@pcl.ac.cn}

\author{Zibin Zheng}
\authornote{Corresponding Author}
\orcid{0000-0002-7878-4330}
\affiliation{%
  \institution{Sun Yat-sen University}
  \city{Zhuhai}
  \country{China}
}
\affiliation{%
  \institution{Guangdong Engineering Technology Research Center of Blockchain}
  \city{Zhuhai}
  \country{China}
}
\email{zhzibin@mail.sysu.edu.cn}

\renewcommand{\shortauthors}{Zexu et al.}

\begin{abstract}
As the development of \textit{Solidity} contracts on \textit{Ethereum}, more developers are reusing them on other compatible blockchains. However, developers may overlook the differences between the designs of the blockchain system, such as the \textit{Gas Mechanism} and \textit{Consensus Protocol}, leading to the same contracts on different blockchains not being able to achieve consistent execution as on \textit{Ethereum}. This inconsistency reveals design flaws in reused contracts, exposing code smells that hinder code reusability, and we define this inconsistency as \textit{EVM-Inequivalent Code Smells}. 

In this paper, we conducted the first empirical study to reveal the causes and characteristics of \textit{EVM-Inequivalent Code Smells}. To ensure the identified smells reflect real developer concerns, we collected and analyzed 1,379 security audit reports and 326 \textit{Stack Overflow} posts related to reused contracts on EVM-compatible blockchains, such as \textit{Binance Smart Chain} (BSC) and \textit{Polygon}. Using the \textit{open card sorting} method, we defined six types of \textit{EVM-Inequivalent Code Smells}. For automated detection, we developed a tool named \textit{EquivGuard}. It employs static taint analysis to identify key paths from different patterns and uses symbolic execution to verify path reachability. Our analysis of 905,948 contracts across six major blockchains shows that \textit{EVM-Inequivalent Code Smells} are widespread, with an average prevalence of 17.70\%. While contracts with code smells do not necessarily lead to financial loss and attacks, their high prevalence and significant asset management underscore the potential threats of reusing these smelly \textit{Ethereum} contracts. Thus, developers are advised to abandon \textit{Copy-and-Paste} programming practices and detect \textit{EVM-Inequivalent Code Smells} before reusing \textit{Ethereum} contracts.

\end{abstract}



\maketitle

\section{Introduction}
\label{sec:intro}
As one of the most active blockchains, \textit{Ethereum} provides a platform for the development and deployment of contracts using the \textit{Ethereum Virtual Machine} (EVM)~\cite{wang2018overview,zou2019smart}. Currently, \textit{Ethereum} has deployed over 65 million contracts and provided a comprehensive ecosystem for developers. To enable better reusability of contracts on \textit{Ethereum}, more than 60\% of blockchains support the execution of EVM, such as \textit{Binance Smart Chain} (BSC)~\cite{bsc}, \textit{Polygon}~\cite{polygon}, and \textit{Arbitrum}~\cite{arbitrum}, which have already captured a significant market share~\cite{evm_compatible,jia2022evm_compat}.

However, developers may overlook the differences between blockchains when reusing contracts, which can even lead to financial loss. Specifically, many blockchains have unique local running environments and settings that differ from \textit{Ethereum}. These differences may result in the same contract code producing inconsistent executions across various EVM-compatible blockchains. For example, \textit{Chain ID}~\cite{wood2014ethereum} uniquely identifies a blockchain network, used to distinguish it from others; directly redeploying \textit{Ethereum} contracts without considering these differences can lead to security issues, as seen in the \$20 million OP token loss by \textit{Wintermute}'s wallet~\cite{OP_replay}. The lack of \textit{Chain ID} validation in \textit{Wintermute}'s \textit{Ethereum} contract allowed attackers to exploit the reused contract on \textit{Optimism} blockchain through signature replay attacks, eventually resulting in huge financial loss.

Thus, it is important to ensure the reusability and reliability when resuing smart contracts to other blockchains. In software engineering, \textit{code smells} are characteristics in the source code that may indicate deeper issues, representing design flaws in the program~\cite{codesmell}. These flaws can impact code reusability, understandability, and reliability.  The inconsistent execution of reused contracts, due to design flaws leading to low reusability and reliability, is becoming increasingly problematic. 

This work is the first to study the inconsistency problem from reused \textit{Ethereum} contracts, termed as \textit{EVM-Inequivalent Code Smells}. To ensure the identified smells reflect real developer issues and guide secure contract reuse, we conducted an empirical study to uncover the causes and characteristics of \textit{EVM-Inequivalent Code Smells}. Specifically, we analyzed security audit reports and \textit{Stack Overflow} posts to collect security issues related to reused contracts on EVM-compatible blockchains, such as \textit{Binance Smart Chain} (BSC)~\cite{bsc} and \textit{Polygon}~\cite{polygon}. Through extensive data collection, we gathered 1,379 security audit reports and 326 relevant posts. Using the \textit{open card sorting} method~\cite{card_sorting}, we identified six types of \textit{EVM-Inequivalent Code Smells}: \textit{Cross-Chain Replay Attack} (CCRA), \textit{Time Discrepancy Trap} (TDT), \textit{Fixed Gas Reentrancy} (FGR), \textit{Block Height Misalignment} (BHM), \textit{Phishing Contract Address} (PCA), and \textit{Gas Limit Imbalance} (GLI) (details in Section~\ref{subsec:defination_examp}). These code smells hinder the reusability and introduce critical security risks for multi-chain reuse contracts.

To investigate the prevalence of the defined six \textit{EVM-Inequivalent Code Smells}, we design an automated detection tool named \textit{EquivGuard}, which utilizes static taint analysis to trace key paths of tainted variable propagation, then employs symbolic execution to verify path reachability and reports detection results. To analyze different code smells, \textit{EquivGuard} identifies key paths using \textit{Domain-Specific Patterns} that encapsulate domain knowledge for identification (details in Section~\ref{subsec:detect_defects}). Specifically, it uses these patterns to pinpoint tainted sources and sinks, analyze taint propagation through global state dependencies, and verify processing safety before reaching sinks. Based on key paths, \textit{EquivGuard} collects path constraints and uses the \textit{SMT} constraint solver to enable path property verification and report detection results.

As \textit{EVM compatibility} has been adopted by many blockchains, we have applied \textit{EquivGuard} to a dataset of 905,948 contracts across six mainstream blockchains, including \textit{Ethereum}, \textit{BSC}, \textit{Arbitrum}, \textit{Polygon}, \textit{Optimism}, and \textit{Avalanche}. Our analysis revealed that 17.70\% of the contracts contained at least one \textit{EVM-Inequivalent Code Smell}, indicating the widespread presence of such code smells in the real world. To evaluate \textit{EquivGuard}'s detection performance for \textit{EVM-Inequivalent Code Smells}, we randomly sampled 444 positive labels and 96 negative labels from the detection results. Through manual analysis, \textit{EquivGuard} achieved an overall precision of 95.29\% and a recall of 95.83\%. Although contracts with code smells do not necessarily lead to financial loss and attacks, their high prevalence and significant asset management highlight the potential threats associated with reusing these smelly \textit{Ethereum} contracts.

The main contributions of this work are as follows:
\begin{itemize}
    \vspace{-0.1cm}
    \item To the best of our knowledge, this study represents the first investigation of \textit{EVM-Inequivalent Code Smells}. We defined six types of code smells by manually analyzing real-world security audit reports and \textit{Stack Overflow} posts, providing definitions and examples for each smell.
    \item We designed \textit{EquivGuard}, combining static taint analysis and symbolic execution. \textit{EquivGuard} achieves an overall 95.29\% detection precision, offering an effective solution for the detection of \textit{EVM-Inequivalent Code Smells}.
    \item We provide a dataset of real-world contracts containing \textit{EVM-Inequivalent Code Smells}. By detecting 905,948 contracts, we found that 17.70\% of the contract source codes contained at least one such code smell. While not all affected contracts necessarily lead to financial loss or attacks, this research encourages developers to recognize these smells and prioritize comprehensive testing across varied environments before reuse, mitigating risks from EVM-inequivalent execution.
\end{itemize}

\section{Background}
\label{sec:background}
\subsection{Explanations of Terminologies}

\textbf{\textcolor{black}{$\blacksquare$} EOAs and Contract Accounts.} \textit{Ethereum} has two account types: External Owned Accounts (EOAs), controlled by private keys, and \textit{Contract Accounts}, whose permissions are determined by their code logic~\cite{2accounts}. While EOAs use the same private key across EVM-compatible blockchains, \textit{Contract Accounts} require separate permission setup on each blockchain, potentially leading to permission loss if not properly configured.

\noindent\textbf{\textcolor{black}{$\blacksquare$} Address.} Address is an identifier for EOA or \textit{Contract Account} locations on the blockchain~\cite{wood2014ethereum}. EOA's addresses are generated by applying the elliptic curve algorithm to a public key and then hashing. \textit{Contract Accounts'} addresses are generated by concatenating the creator's address with the creation transaction's nonce and then hashing the result. This means that a contract address may deploy different code on EVM-compatible blockchains. For example, the contract code at address 0x818ec0a7~\footnote{0x818ec0a7fe18ff94269904fced6ae3dae6d6dc0b} differs between \textit{Ethereum} and \textit{Moonbeam}.


\noindent\textbf{\textcolor{black}{$\blacksquare$} EIPs.} 
Ethereum Improvement Proposals (EIPs), like \textit{EIP-2612} which introduced the \textit{permit()} function for offline signature validation~\cite{eip2612}, aim to change or update \textit{Ethereum} by enhancing its capabilities and security features. However, many EVM-compatible blockchains do not fully comply with EIPs, potentially impacting their EVM compatibility and leading to inconsistencies in how certain features and functions are implemented across different platforms. This lack of standardization can pose challenges for developers aiming for cross-chain interoperability and uniformity in smart contract behavior.

\noindent\textbf{\textcolor{black}{$\blacksquare$} Gas Mechanism.} Gas measures the computational effort on the blockchain. Gas cost quantifies the amount of gas required to execute each operation, e.g., \textit{Ethereum}'s \textit{ADD} operation consumes 3 gas, according to the Gas cost standard~\cite{evmcode}. This standard continuously evolves, for instance, EIP-1380~\cite{eip1380} reduced the Gas cost for the call to self from 700 to 40. Additionally, many EVM-compatible blockchains have their own Gas cost standards and refund mechanisms to offer lower transaction fees, such as BSC's BEP~\cite{bep}.

\noindent\textbf{\textcolor{black}{$\blacksquare$} Transfer()/Send().} \textit{Transfer()} and \textit{Send()} functions are used for token transfers, with a fixed 2300 gas limit commonly used to prevent Reentrancy attacks. However, this security measure depends on the Gas cost remaining unchanged~\cite{transfersend}.

\noindent\textbf{\textcolor{black}{$\blacksquare$} Hard Fork.} Hard Fork significantly changes blockchain protocols, resulting in a split that creates a new blockchain~\cite{hardfork}. This process is often used to implement major changes, leading to two versions of the blockchain with distinct paths. This split allows the new protocol to integrate advanced features or security improvements, though it requires consensus to avoid fragmentation.

\noindent\textbf{\textcolor{black}{$\blacksquare$} Block Time.} Block time is the average time it takes for a new block to be added to a blockchain~\cite{BCT}.

\subsection{EVM-compatible Blockchains}

EVM-compatible blockchains support EVM and \textit{Solidity} contracts, enabling developers and users to build DApps across multiple blockchains and reducing barriers to contract deployment and interaction~\cite{evm_compatible}. Table~\ref{tab:bg1} presents the statistics of EVM-compatible blockchain assets on \textit{DefiLlama}. Out of the total, 150 are EVM-compatible blockchains, accounting for 63.03\%. The \textit{Total Value Locked} (TVL) of these EVM-compatible blockchains reached \$90.997 billion, significantly surpassing the \$14.213 billion of non-EVM-compatible blockchains. These findings show that EVM-compatible blockchains hold over 86\% of the market share, underscoring the importance of EVM compatibility for blockchain networks.

\begin{table}[htb]
    \centering
    \caption{Distribution of Chains and Total Value Locked (TVL)}
\begin{tabular}{l|llll}
\hline
                            & \textbf{Chains} & \textbf{Percentage} & \textbf{TVL (billion)} & \textbf{TVL Percentage} \\ \hline
\textbf{EVM-compatible}     & 150             & 63.03\%             & 90.997                & 86.49\%                 \\
\textbf{Non-EVM-compatible} & 88              & 36.97\%             & 14.213                 & 13.51\%                  \\ \hline
\end{tabular}
    \label{tab:bg1}
\end{table}

We compared the fees and speeds of transactions on the top 5 EVM-compatible blockchains by market share with \textit{Ethereum} (average data from September 2024). As shown in Table~\ref{tab:bg2}, BSC's \textit{Time To Finality} (TTF) is only 7.5s, and the average transaction fee on \textit{Polygon} is minimal, at \$0.013. While EVM-compatible blockchains offer lower transaction fees or faster speeds, many blockchain projects must restructure their native virtual machine and continuously update improvement proposals to achieve EVM compatibility. Most blockchains only partially follow or do not follow \textit{Ethereum Improvement Proposals} (EIPs), with only \textit{Polygon} fully supporting them. These modifications and inconsistent implementations hinder the full adaptation of the EVM module on these blockchains, affecting EVM compatibility.

\begin{table}[htb]
    \centering
    \caption{Statistics of Top 5 EVM-Compatible Blockchains}
        \begin{tabular}{lllll}
            \hline
            \textbf{Blockchain} & \textbf{Txn Fee} & \textbf{Block Time} & \textbf{TTF} & \textbf{EIP-compliant} \\
            \hline
            \ {\includegraphics[width=4mm, height=3mm]{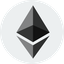} \textit{Ethereum}} & \$15.59          & 12.14s              & 16m          & Fully \\
            \ {\includegraphics[width=3.5mm, height=2.8mm]{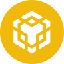} \textit{BSC}}                                    & \$0.24           & 3.01s               & 7.5s         & Partial \\
            \ {\includegraphics[width=3.5mm, height=2.8mm]{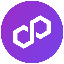} \textit{Polygon}}                                & \$0.013          & 2.26s               & 4m16s        & Fully \\
            \ {\includegraphics[width=3.5mm, height=2.8mm]{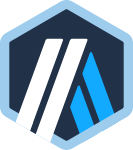} \textit{Arbitrum}}                               & \$0.28           & 0.26s               & 16m          & Partial \\
            \ {\includegraphics[width=3.5mm, height=2.8mm]{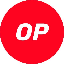} \textit{Optimism}}                               & \$0.069          & 2s                  & 16m          & Partial \\
            \ {\includegraphics[width=3.5mm, height=2.8mm]{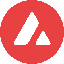} \textit{Avalanche}}                              & \$0.23           & 2.04s               & 0s           & Not \\
            \hline
        \end{tabular}  
    \label{tab:bg2}
\vspace{0.1cm}
\textsuperscript{\small\textit{*\textbf{TTF} refers to the time it takes for a transaction to be confirmed.}}
\end{table}  

\section{EVM-Inequivalent Code Smells}
\label{sec:defects}
In this section, we conduct an empirical study on real-world security audit reports and \textit{Stack Overflow} posts related to EVM-compatible blockchains to define and classify \textit{EVM-Inequivalent Code Smells} in reuse contracts.

\subsection{Data Collection}
\label{sec:dadacollect}
To comprehensively analyze real-world \textit{EVM-Inequivalent Code Smells}, we collected 1,379 publicly available audit reports from 30 security teams and 326 relevant \textit{Stack Overflow} posts. We then conducted a thorough manual screening to filter the relevant information for further analysis. 

\subsubsection{Security Audit Reports Collection}
The \textit{Security Audit Report} (SAR) contains specific vulnerabilities and cause analyses, providing a comprehensive understanding of security issues. To collect audit reports related to \textit{EVM-Inequivalent Code Smells} in the real world, we accessed the websites of 81 smart contract security teams listed on \textit{Etherscan}~\cite{etherscanaudit} and identified 30 teams that had open-sourced their audit reports, including \textit{SlowMist}~\cite{slowmist}, \textit{ConsenSys}~\cite{consensys}, \textit{BlockSec}~\cite{blocksec}, and \textit{Trail of Bits}~\cite{TrailofBits}. Additionally, based on non-zero \textit{Total Value Locked} (TVL) and using Solidity contracts, we filtered out 150 EVM-related blockchains from the 207 blockchains listed on \textit{DefiLlama}. Combining this with audit report public websites~\cite{solodit}, we collected a total of 1,379 EVM-related audit reports.

\subsubsection{Stack Overflow Posts Collection}
\textit{Stack Overflow}~\cite{stackoveflow} is a popular Q\&A community for developers, engineers, and technology enthusiasts. The platform uses a tag system to manage topics involved in the questions, allowing us to collect the issues and concerns encountered by developers quickly. To ensure efficient analysis of \textit{Stack Overflow Posts (SOP)} related to EVM inequivalence, we initially collected 2,124 posts using the tags ``\textit{Solidity Contract}", ``\textit{EVM}" to gather posts related to \textit{Solidity contracts} on \textit{EVM chains}. Subsequently, to filter out irrelevant information and balance human efforts, we used the keywords ``\textit{EVM Equivalent}" and ``\textit{EVM Compatible}" for further screening, resulting in 326 related posts on contract reuse in EVM.

\subsubsection{Manual Screening} To filter relevant information, we manually analyzed the collected reports and posts. We assigned two authors with extensive smart contract development experience to manually filter through 1,379 audit reports and 326 related posts, retaining only content related to \textit{EVM-Inequivalent Code Smells}. \textbf{\textit{For audit report filtering.}} Audit reports include \textit{Issues}, \textit{Root Causes}, and related \textit{Recommendations}. Two authors quickly determined whether the issues were related to EVM-Inequivalent caused by contract reuse by reading these sections. \textbf{\textit{For Stack Overflow post filtering.}} Stack Overflow posts also include \textit{Titles}, \textit{Descriptions}, and \textit{Comments}. Two authors quickly identified and removed irrelevant content. Additionally, if the authors could not directly determine whether a post should be retained based on its content, it was categorized as "\textit{tentative}". Finally, the authors compared results and discussed differences, identifying 197 relevant security analyses of contract reuse on EVM-compatible blockchains.

\subsection{Data Analysis}
\label{sec:dadaanalysis}
To classify \textit{EVM-Inequivalent Code Smells}, we utilized \textit{Card Sorting}, a user-involved design method, to organize and categorize them without predefined rules~\cite{cardsorting1}. For each security report or \textit{Stack Overflow} post, we created a card with different sections, allowing readers to sort and filter based on their understanding and preferences.

\begin{figure}[h]
    \centering
    \includegraphics[width=0.65\columnwidth]{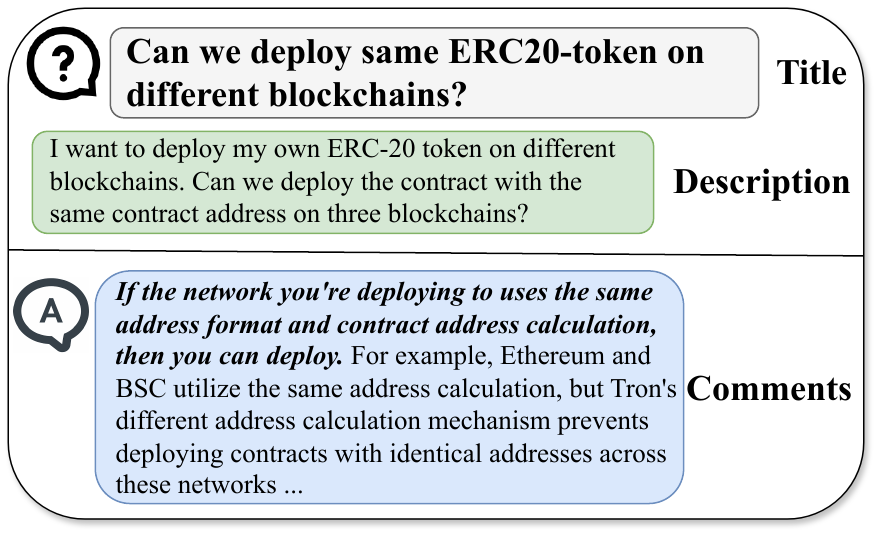}
    \caption{An example of the \textit{Stack Overflow} Post (SOP).}
    \label{fig:post_example}
\end{figure}

Figure~\ref{fig:post_example} shows an example of the \textit{Stack Overflow} post card, divided into three parts: \textit{Title}, \textit{Description}, and \textit{Comments}. The post's author wants to deploy ERC-20 tokens using the same contract address on multiple blockchains and seeks a solution. Replies in the \textit{Comments} mention the possibility of deploying the contract with the same address format and deployment method on \textit{Ethereum}~\cite{ethereum} and \textit{Binance Smart Chain} (BSC)~\cite{bsc}, but \textit{Tron} network~\cite{tron} generates addresses differently, making it impossible to deploy a contract with the same address as \textit{Ethereum}. This highlights that even the same contract cannot guarantee the same address across different blockchains due to the varying address generation methods~\cite{SOpost, evmaddress}. The post, viewed 3,000 times, highlights the significance of this issue: reused contracts containing calls to fixed-address contracts can be easily exploited for phishing scams on EVM-compatible blockchains. This problem is representative, reproducible, and worthy of attention, summarized as \textit{Phishing Contract Address} (PCA).

\begin{figure}[htb]
    \centering
    \includegraphics[width=0.65\columnwidth]{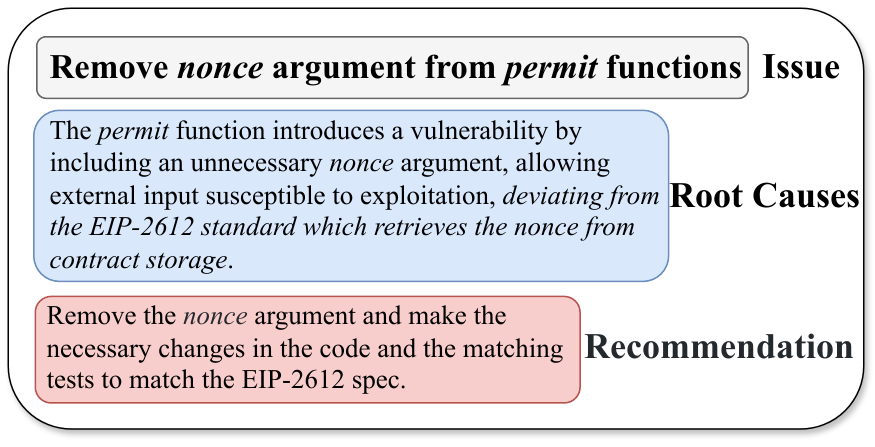}
    \caption{An example of the Security Audit Report (SAR).}
    \label{fig:report}
\end{figure}

Figure~\ref{fig:report} shows an example of the security audit report card, including three parts: \textit{Issue}, \textit{Root Causes} and \textit{Recommendation}. The report pointed out that the \textit{permit()} function did not strictly follow the \textit{EIP-2612} standard~\cite{eip2612}, and there was an extra \textit{nonce} parameter in the input parameter. As the \textit{nonce} cannot be specified from external input, the auditors recommended removing the \textit{nonce} parameter in the \textit{permit()} function and removing the statement that verifies the \textit{nonce} to ensure consistency with the \textit{EIP-2612} standard. After discussion, we classified this issue as a \textit{Cross-Chain Replay Attack} (CCRA) due to the incorrect signature verification, which could lead to cross-chain replay attacks, particularly when reusing contracts across different EVM-compatible blockchains.

Two authors analyzed 197 collected cards to classify code bad smells in reuse contracts. They examined the root causes, descriptions and comments, placing code smells into existing categories if possible. If not, they analyzed whether the problems were representative to determine if new categories were needed. The authors compared results, discussed differences, and identified six types of \textit{EVM-Inequivalent Code Smells}.

\begin{table}[htb]
    \centering
    \caption{Statistical Distribution of Different Code Smells Sources}
\begin{tabular}{l|llllll}
\hline
\textbf{Type}       & \textbf{CCRA} & \textbf{TDT} & \textbf{PCA} & \textbf{GLI} & \textbf{FGR} & \textbf{BHM} \\ \hline
\textbf{Quantity}   & 82            & 32           & 11           & 38           & 13           & 21           \\
\textbf{Percentage} & 41.62\%       & 16.24\%      & 5.58\%       & 19.29\%      & 6.60\%       & 10.66\%      \\
\textbf{SOP/SAR}    & 0/82          & 15/17        & 11/0         & 0/38         & 0/13         & 0/21         \\ \hline
\end{tabular}
    \label{tab:statist}
\end{table}

Table~\ref{tab:statist} provides statistics on the sources of different defect types, showcasing their distribution in \textit{Stack Overflow Posts} (SOPs) and \textit{Security Audit Reports} (SARs). \textit{Cross-Chain Replay Attack} (CCRA) type leads with 82 instances, accounting for 41.62\% of the total. \textit{Gas Limit Imbalance} (GLI), \textit{Fixed Gas Reentrancy} (FGR), and \textit{Block Height Misalignment} (BHM) defects are only recorded in SARs, suggesting these problems are more known to professional auditors and less known to the external developer community. In contrast, \textit{Phishing Contract Address} (PCA) defects are only noted in SOPs, indicating these are problems developers encounter and seek help for. \textit{Time Discrepancy Trap} (TDT) defects are collected with 15 SOPs and 17 SARs, highlighting this defect type as a focus in the developer community and in audits. These data reveal differences in the sources of defect types, helping us understand the common issues developers face when reusing contract codes.

\subsection{\textit{EVM-Inequivalent Code Smells} Definition}
\label{subsec:defination_examp}
In this section, we provide definitions of six types of \textit{EVM-Inequivalent Code Smells} in Table~\ref{tab:types_of_defects}, followed by comprehensive definitions and illustrative code samples.

\begin{table*}[h]
    \centering
    \large
    \caption{Definitions of \textit{EVM-Inequivalent Code Smells} in \textit{Ethereum}’s reuse contracts}
    \resizebox{0.98\linewidth}{!}{%
\begin{tabular}{l|l|l}
\hline
\textbf{Code   Smell} & \textbf{ID} & \textbf{Definition}                                                           \\ \hline
\textit{Cross-Chain Replay Attack}  & CCRA        & Incorrect chain information validation leads to cross-chain transaction replay.    \\
\textit{Time Discrepancy Trap}      & TDT         & The changing block time causes unexpected state updates.        \\
\textit{Fixed Gas Reentrancy}       & FGR         & Reentrancy vulnerabilities caused by changes in Gas cost.                    \\
\textit{Block Height Misalignment}  & BHM         & Different block heights cause inconsistent execution.            \\
\textit{Phishing Contract Address}  & PCA         & Stealing permissions from a designated contract address and impersonating it. \\
\textit{Gas Limit Imbalance}        & GLI         & The specified gas limit in the contract leads to denial of service.            \\ \hline
\end{tabular}
    }
    \label{tab:types_of_defects}
\end{table*}

 \textbf{(1) \textit{Cross-Chain Replay Attack} (CCRA).} Signatures allow users to authorize others to execute transactions and ensure their legitimacy. Signature verification typically involves validating chain-specific information, such as the \textit{Chain ID}, to prevent signature replay across different blockchains~\cite{eip2612}. However, due to a lack of security awareness among developers, many multi-chain contracts lack chain-specific checks during signature verification, leaving them vulnerable. When these contracts are reused, they become vulnerable to replay attacks, where attackers can obtain signatures on \textit{Ethereum} and replay them on other EVM-compatible blockchains.

\textbf{Code Example:} In Figure~\ref{MExample3}, \textit{verifyEIP712()} function (line 7) validates the signature information \textit{(v, r, s)} to confirm the signer's authenticity. It utilizes the \textit{ecrecover()} to recover the signer's address and checks if the address matches the target address~\cite{ecrecover}. The problem stems from the \textit{setter()} function (line 3), allowing anyone to change the \textit{chainId} and \textit{DOMAIN\_SEPARATOR} variables, making permission verification easy to bypass. Attackers can exploit the lack of \textit{Chain ID} verification to perform cross-chain signature replay attacks. 

 \vspace{-0.1cm}
\begin{figure}[htb]
	\setlength{\abovecaptionskip}{0.2cm}
	\begin{lstlisting}[xleftmargin=0.1cm,xrightmargin=0.2cm]
 bytes32 public DOMAIN_SEPARATOR;
 uint256 public chainId;
 function setter(uint256 _chainId) public {
     chainId = _chainId;
     DOMAIN_SEPARATOR = keccak256(abi.encode(keccak256(..., chainId,address(this))));
 }
 function verifyEIP712(address target, bytes32 hashStruct, uint8 v, bytes32 r, bytes32 s) public view returns (bool) {
     bytes32 hash = keccak256(abi.encodePacked("\x19\x01", DOMAIN_SEPARATOR, hashStruct));
     address signer = ecrecover(hash, v, r, s);
     return(signer == target);
 }
	\end{lstlisting}
	\caption{The example of \textit{Cross-Chain Replay Attack} (CCRA)}
	\label{MExample3}
\end{figure}
\vspace{-0.1cm}

 \textbf{(2) \textit{Time Discrepancy Trap} (TDT).} Developers prefer using \textit{block.number} over \textit{block.timestamp} to calculate time intervals, as it is less susceptible to miner manipulation attacks~\cite{TDTcase}. However, this can lead to \textit{Time Discrepancy Trap} (TDT) when reusing contracts across multiple blockchains due to varying block generation times (e.g., approximately 15 seconds on \textit{Ethereum}).  Failing to consider these differences in block generation times when reusing contracts can lead to unexpected results.

 \vspace{-0.2cm}
\begin{figure}[htb]
	\begin{lstlisting}[xleftmargin=0.1cm,xrightmargin=0.2cm]
 uint256 constant public BLOCKS_PER_WEEK = 43200; // The approximate number of Ethereum blocks per week.
 function TimeLockedWithdraw() external {
     require(block.number >= depositBlock[msg.sender] + BLOCKS_PER_WEEK, "Funds are still locked!");
     balances[msg.sender] = 0;
     payable(msg.sender).transfer(balances[msg.sender]);
 }
	\end{lstlisting}
	\caption{The example of \textit{Time Discrepancy Trap} (TDT)}
	\label{MExample5}
\end{figure}
\vspace{-0.2cm}

\textbf{Code Example:} In Figure~\ref{MExample5}, the \textit{TimeLockedWithdraw()} function (lines 2--6) calculates time intervals using \textit{BLOCKS\_PER\_WEEK}, reducing reliance on timestamps and improving cross-chain deployment. However, on EVM-compatible blockchains with faster block generation, the 43,200 block time might not accurately represent a week, impacting calculation accuracy. This inconsistency, due to varying block generation times, can cause the actual withdrawal time to deviate from the intended 1-week, impacting the logic's reliability and predictability. Attackers could exploit this by triggering early or delayed withdrawals, potentially leading to fund loss or logic violations. 

 \textbf{(3) \textit{Phishing Contract Address} (PCA).} To enhance code readability and simplify development, developers often use fixed addresses in contract, such as for Uniswap routers~\cite{uniswaprouter}, a common practice in certain industries. However, when reusing contracts across multiple chains, these hardcoded addresses can lead \textit{Phishing Contract Address} (PCA), which attackers can exploit. This vulnerability arises because \textit{Contract Account}'s ownership is determined by contract logic, preventing direct ownership transfer across blockchains. Attackers can leverage the \textit{create()} function to deploy contracts and gain ownership of the generated address on different chains. As the \textit{Wintermute} attack, attackers exploited signature replay and nonce collision techniques to take over target addresses on Optimism~\cite{OP_replay}, impersonating the victim and launching phishing attacks.

 \vspace{-0.2cm}
\begin{figure}[htb]
	\setlength{\abovecaptionskip}{0.2cm}
	\begin{lstlisting}[xleftmargin=0.1cm,xrightmargin=0.2cm]
 contract SwapToken {
    // Uniswap Router address on mainnet
    address RouterAddr = 0x7a250d56...59F2488D; 
    RouterV2 uniswapRouter = RouterV2(RouterAddr);
    // Exchange ETH for Token in Uniswap
    function swapEthForTokens(...) external payable {
        uniswapRouter.swapExactETHForTokens(...);
     }
 }
	\end{lstlisting}
	\caption{The example of \textit{Phishing Contract Address} (PCA)}
	\label{MExample1}
\end{figure}
\vspace{-0.2cm}

\textbf{Code Example:} Figure~\ref{MExample1} shows a contract utilizing \textit{Uniswap}'s Router contract address~\cite{uniswaprouter} on the \textit{Ethereum} mainnet. While \textit{Uniswap}'s official EOA has permissions on \textit{Ethereum}, these permissions are unknown or undefined on other EVM-compatible blockchains. This opens a window for attackers to deploy a fake Router contract on these alternative chains and exploit the \textit{swapEthForTokens()} function (line 6) to steal ETH and distribute counterfeit tokens. This vulnerability exists because contract ownership cannot be directly transferred across blockchains in the same way as EOAs. As a result, attackers can deploy contracts with identical addresses on different blockchains, enabling sophisticated phishing attacks that exploit these discrepancies.

 \textbf{(4) \textit{Gas limit Imbalance} (GLI).} Gas measures the computational effort needed to execute contracts, preventing blockchain resource abuse. Developers often set a fixed gas limit in contracts. However, setting an excessively high or low gas limit can lead to issues~\cite{GasLimit}. Due to varying Gas cost standards across different blockchains, a unified gas limit can be easily exploited by attackers. When reusing contracts across blockchains, developers need to dynamically adjust gas limits to ensure successful transactions and avoid wasting gas.
 
\vspace{-0.1cm}
\begin{figure}[htb]
	\setlength{\abovecaptionskip}{0.2cm}
	\begin{lstlisting}[xleftmargin=0.1cm,xrightmargin=0.2cm]
 Payee[500] payees;
 uint256 nextPayeeIndex;
 function payOut() public returns (uint) {
     uint256 i = nextPayeeIndex;
     while (i < payees.length && gasleft() > 400000) {
         payees[i].value = 0;
         payees[i].value=payees[i].value+1;
         i++;
     }
     nextPayeeIndex = i;
    return nextPayeeIndex;
 }
	\end{lstlisting}
	\caption{The example of Gas limit imbalance (GLI)}
	\label{MExample2}
\end{figure}
\vspace{-0.1cm}

\textbf{Code Example:} The \textit{payOut()} function (lines 3--12) in Figure~\ref{MExample2} iterates over the \textit{Payee[]} array, which has a length of 500. It uses the \textit{gasleft()} function (line 5) to ensure that the current remaining gas is greater than 400,000. However, when reusing this code on blockchains with lower gas cost standards, the 400,000 gas requirement can easily cause transactions to fail due to the high limit. This results in the loop exiting prematurely, preventing some elements of the \textit{Payee[]} array from being processed normally and potentially triggering a \textit{Denial of Service} (DoS) attack. This issue underscores the importance of adjusting gas limits to accommodate different blockchain environments, ensuring reliable execution.

 \textbf{(5) \textit{Fixed Gas Reentrancy} (FGR).} The \textit{transfer() }and \textit{send()} functions can effectively prevent Reentrancy attacks on \textit{Ethereum} by limiting the maximum Gas consumption to 2300 units. This limitation restricts the ability to execute additional function calls or alter contract states within the scope of a transaction on \textit{Ethereum}. However, this mechanism may not be consistently implemented across different blockchains, or other platforms may feature lower Gas cost standards~\cite{transfersend}. Consequently, when a contract is redeployed on an EVM-compatible blockchain without adjusting for these discrepancies, it may become susceptible to security vulnerabilities. 

\vspace{-0.1cm}
\begin{figure}[htb]
	\setlength{\abovecaptionskip}{0.2cm}
	\begin{lstlisting}[xleftmargin=0.1cm,xrightmargin=0.2cm]
 function withdraw() external {
    // transfer(): Send funds, 2300 gas fixed.
    payable(msg.sender).transfer(balances[msg.sender]);
    // Update user balance
    balances[msg.sender] = 0;
 }
	\end{lstlisting}
	\caption{The example of \textit{Fixed Gas Reentrancy} (FGR)}
	\label{MExample4}
\end{figure}
\vspace{-0.1cm}

\textbf{Code Example:} As shown in the \textit{withdraw()} function in Figure~\ref{MExample4}, the caller is allowed to transfer money through \textit{transfer()}. However, if the caller is a malicious contract, its fallback function may call the \textit{withdraw()} function again when receiving Ether, completing a Reentrancy attack. Although the 2300 gas limit can currently prevent Reentrancy attacks on \textit{Ethereum}, this defense may be ineffective on other blockchains with lower Gas cost standards. To achieve lower transaction fees, the Gas cost of many EVM-compatible blockchains is constantly changing. Therefore, relying on a specific 2300 Gas costs is a vulnerable pattern that cannot fundamentally eliminate the occurrence of Reentrancy vulnerabilities~\cite{transfersend}.

 \textbf{(6) \textit{Block Height Misalignment} (BHM).} 
Block height, being a relatively stable metric, is often used by developers in smart contracts for operations like voting, governance, and contract upgrades. However, when reusing these contracts on different EVM-compatible blockchains, each blockchain has its independent block height progression. The specific heights referenced in the code may not match the target blockchain's heights. This mismatch can lead to severe consequences, such as the contract logic failing to execute intended operations at desired block heights or causing unintended consequences due to differences in block height progression between blockchains~\cite{Blockheigth}. Attackers can exploit this, potentially leading to fund loss, governance issues, or critical failures.

\vspace{-0.2cm}
\begin{figure}[htb]
	\setlength{\abovecaptionskip}{0.2cm}
	\begin{lstlisting}[xleftmargin=0.1cm,xrightmargin=0.2cm]
 // Hard fork for DAO
 uint constant DAO_FORK_BLOCK = 1760000; 
 function handleFork() public {
     if (block.number >= DAO_FORK_BLOCK) {
     ...
 } 
	\end{lstlisting}
	\caption{The example of \textit{Block Height Misalignment} (BHM)}
	\label{MExample6}
\end{figure}
\vspace{-0.2cm}

\textbf{Code Example:} The \textit{handleFork()} function in Figure~\ref{MExample6} is designed to handle the DAO fork on \textit{Ethereum} by setting a specific block height,  \textit{DAO\_FORK\_BLOCK} (at 1760000), to identify the event's occurrence. The function compares the current block height with the DAO fork's block height to execute the processing logic. An overly large block height requirement could trigger a \textit{Denial of Service} attack, or changes in block height progression may cause unintended executions. Therefore, it is crucial to carefully consider using fixed block heights when reusing smart contract code on different EVM-compatible blockchains to ensure robustness and security.

\section{Methodology}
\label{sec:method}
\subsection{Overview}

\begin{figure}[h]
    \centering
    \includegraphics[width=0.83\columnwidth,height=58mm]{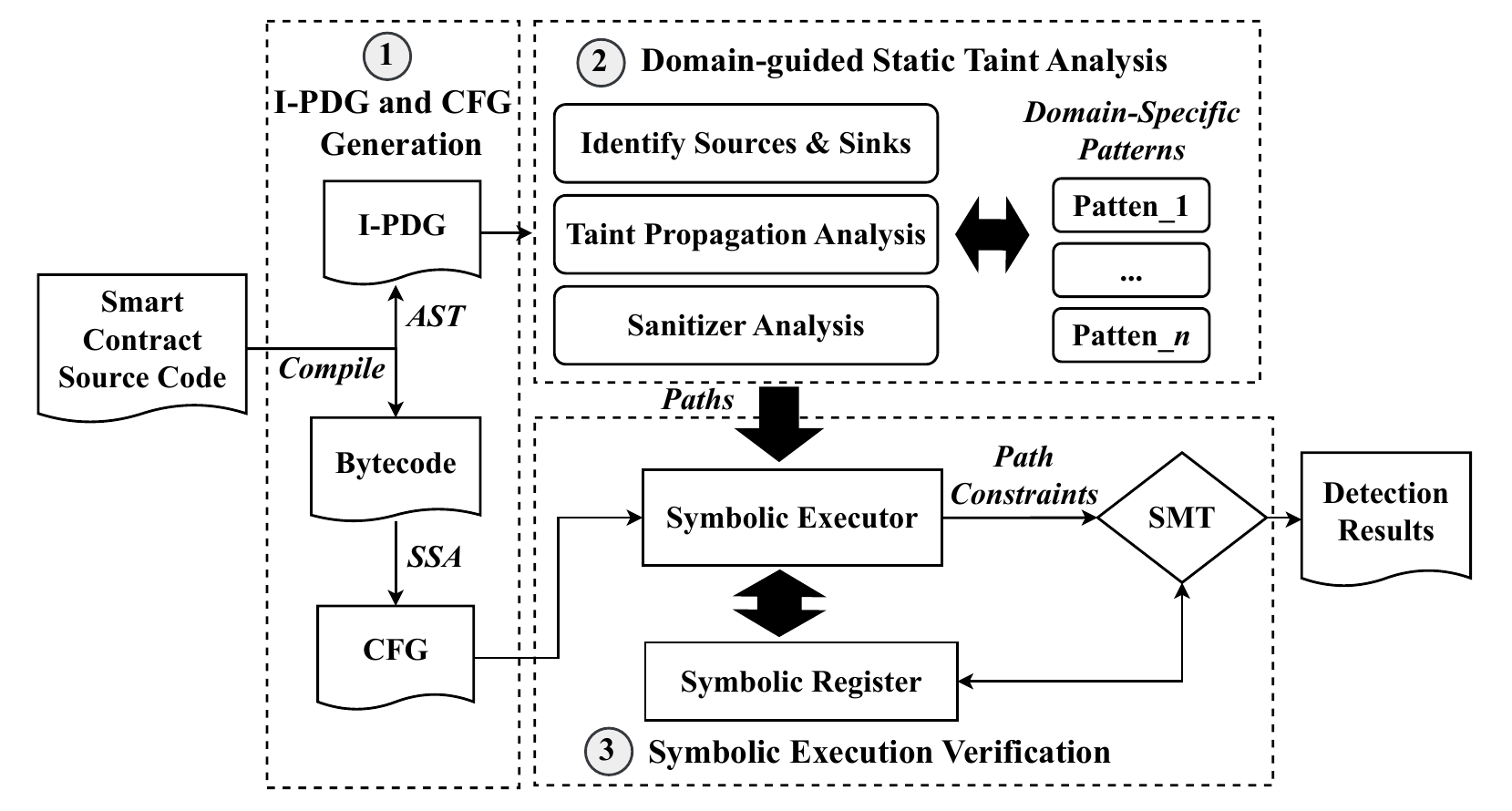}
    \caption{The workflow of \textit{EquivGuard}.}
    \label{fig:overview}
\end{figure}

The workflow of \textit{EquivGuard} is shown in Figure~\ref{fig:overview}. It takes the \textit{Ethereum} Solidity contract source code as input because \textit{Ethereum} contracts are most likely reused by other Blockchains. The detection process consists of three main steps: generating \textit{Inter-contract Program Dependency Graph} (I-PDG)~\cite{wang2024efficiently} and the \textit{Control Flow Graph} (CFG)~\cite{makingsmart}, identifying suspicious paths through static taint analysis, and verifying path feasibility via symbolic execution. \textit{\textbf{Step 1: I-PDG and CFG Generation}}. \textit{EquivGuard} analyzes the contract source code to generate two graphs, which serve as inputs for Step 2 and Step 3. \textit{\textbf{Step 2: Domain-Guided Static Taint Analysis}}. We utilize different \textit{Domain-Specific Patterns} to guide taint propagation analysis, identify and track potentially paths. Combining different exploitation modes, key instructions, and reverse analysis enables efficient path search. \textit{\textbf{Step 3: Symbolic Execution Verification}}. Based on the suspicious path information, \textit{EquivGuard} verifies the reachability of the path through symbolic execution to improve the accuracy of code smell detection. Finally, \textit{EquivGuard} outputs the detection results.

\subsection{I-PDG and CFG Generation}
To model and analyze contract semantics, \textit{EquivGuard} compiles the source code to obtain the \textit{Abstract Syntax Tree} (AST) and bytecode. It then constructs the \textit{Inter-contract Program Dependency Graph} (I-PDG)~\cite{wang2024efficiently} and the \textit{Control Flow Graph} (CFG)~\cite{makingsmart}, which capture program dependencies and control flow, respectively. These graphs serve as inputs for Step 2 and Step 3 of the process. 

The AST provides rich syntactic information about the code, including the positions and relationships of keywords, variables, and instructions. This information aids in the global dependency analysis and construction of the I-PDG, which is generated by integrating program dependency analysis between AST statement blocks into the \textit{Inter-contract Control Flow Graph} (I-CFG)~\cite{ma2021pluto}. The I-PDG contains the global program dependencies of the contract and systematically analyzes the program semantics of smart contracts. With accurate data and control dependencies, the I-PDG facilitates efficient data flow tracking, identification of potential logical errors, and analysis of special semantics. For example, when detecting \textit{Cross-Chain Replay Attack} (CCRA), the \textit{ecrecover()} function is critical. \textit{EquivGuard} uses AST analysis to directly determine the dependency path set from the external function entry to the \textit{ecrecover()} function call instruction, combining static taint analysis between key instructions to identify dangerous vulnerabilities.

While SSA facilitates the analysis of underlying execution details, its lack of semantic context makes it ineffective for the global syntactic analysis of contracts. The CFG generated from bytecode is crucial for symbolic execution path traversal. During CFG path recovery, we convert stack operations in the bytecode to SSA form and perform constant propagation analysis to solve the dynamic jump address calculation problem. By converting the bytecode to \textit{Static Single Assignment} (SSA) form~\cite{wang2024efficiently}, we ensure complete path recovery, enabling the symbolic executor to collect comprehensive path constraints.

\subsection{Domain-guided Static Taint Analysis}
As smart contracts become larger and more complex, involving intricate function calls, contract inheritance, and cross-contract calls, among others, the growth of path states has exploded. Heuristic and coarse-grained data flow analysis can easily lead to overfitting problems, seriously affecting the accuracy of results. To address these challenges, we propose data flow analysis and path reverse analysis based on key instructions to achieve an efficient path search for static taint analysis.

\begin{figure}[htb]
    \centering
    \includegraphics[width=\columnwidth]{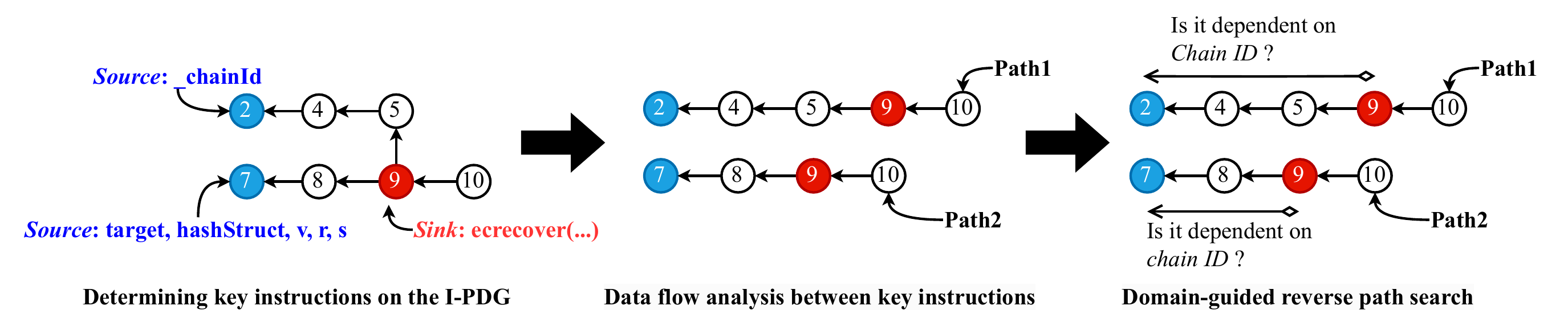}
    \caption{Domain-Guided Static Taint Analysis for \textit{Cross-Chain Replay Attack} (CCRA). The numbers in the circles are the line numbers from Figure~\ref{MExample1}. Blue circles are \textit{Source} points, red circles are \textit{Sink} points, and arrows show program dependencies.}
    \label{fig:case1}
\end{figure}

{\textbf{(1) Data flow analysis between key instructions.}} \textit{EquivGuard} utilizes I-PDG with key instructions to implement data flow analysis between global instructions. The I-PDG converts global function calls and cross-contract calls into code block~\footnote{Code block refers to a sequence of instructions in the contract's control flow that are executed sequentially without branching or function calls.} jumps in the graph, ensuring correct processing of return values. This helps identify patterns and perform sanitizer analysis by revealing dependencies between contract variables, improving static taint analysis accuracy and efficiency. At the same time, \textit{EquivGuard} identifies key instructions as sinks, marks critical variables influenced by external inputs as sources, and performs taint propagation analysis based on the I-PDG while conducting sanitizer analysis. \textit{EquivGuard} avoids traversing irrelevant programs by analyzing paths between key instructions of patterns, thereby improving the accuracy and flexibility of the analysis. The selection of key instructions depends on different \textit{Domain-Specific Patterns}. As shown in Figure~\ref{fig:case1}, for static taint analysis of \textit{Cross-Chain Replay Attack} (CCRA), \textit{EquivGuard} uses the \textit{ecrecover()} call instruction as \textit{Taint} and external inputs as \textit{Source}. By obtaining two different paths from the dependency relationship between the key instructions of \textit{Source} and \textit{Taint}, further data flow analysis is achieved. The \textit{ecrecover()} function is a key function, and \textit{EquivGuard} uses key instruction guidance to ensure efficient path search. For more detailed detection methods, please refer to section~\ref{subsec:detect_defects}.

{\textbf{(2) Domain-guided reverse path search.}}
To prune paths, \textit{EquivGuard} starts from the variable usage statement in the critical function and traces it backward to the assignment or declaration point of the variable. Therefore during static taint analysis, starting from the taint sinks, the data flow is traced back to its sources. Complex function calls and cross-contract calls involved are analyzed through I-PDG, identifying any sanitizer mechanisms in the data flow. Based on the I-PDG's data flow analysis, \textit{EquivGuard} effectively focuses on states and value sources when the taint sinks are used.

Since \textit{EVM-Inequivalent Code Smells} are closely related to the incorrect dependence of the reused contract on the specific blockchain state, the main challenge is to analyze whether the variable's value remains unchanged throughout the execution without running the program code. As shown in Figure~\ref{fig:case1}, \textit{EquivGuard} starts from the taint sinks on different paths and uses dependency relationships to reverse analyze any dynamic dependence on \textit{blockchainID} (from the blockchain environment). \textit{EquivGuard} combines I-PDG with reverse analysis of critical functions to analyze variable propagation and detect smells. This reverse tracing helps analyze any contamination relationship between function input and output and whether the data has been verified, cleaned, or protected, achieving efficient attribute verification of paths.

\subsection{Symbolic Execution Verification}
To improve detection accuracy, \textit{EquivGuard} employs symbolic execution to traverse suspicious paths generated in \textit{Step 2} and verify their feasibility. To mitigate the effect of calling permissions determined by specific rules that may affect path analysis and result in false positives, we introduce symbolic execution to collect path constraints and verify the reachability of suspicious paths, thereby enhancing the accuracy of code smell detection. \textit{EquivGuard} traverses the CFG branch to model the contract's storage and collect path constraints, then proves reachability through path constraint solving. Therefore, we introduce the \textit{Symbolic Register} to store symbolic states and path constraint information, overcoming the challenge of incomplete path constraint collection. By analyzing the state storage operation of bytecode and combining it with the \textit{Z3 solver}, we store specific values and symbolic expressions as key-value pairs in the \textit{Symbolic Register}.

\vspace{-0.3cm}
\begin{figure}[htb]
    \centering
    \includegraphics[width=0.88\columnwidth]{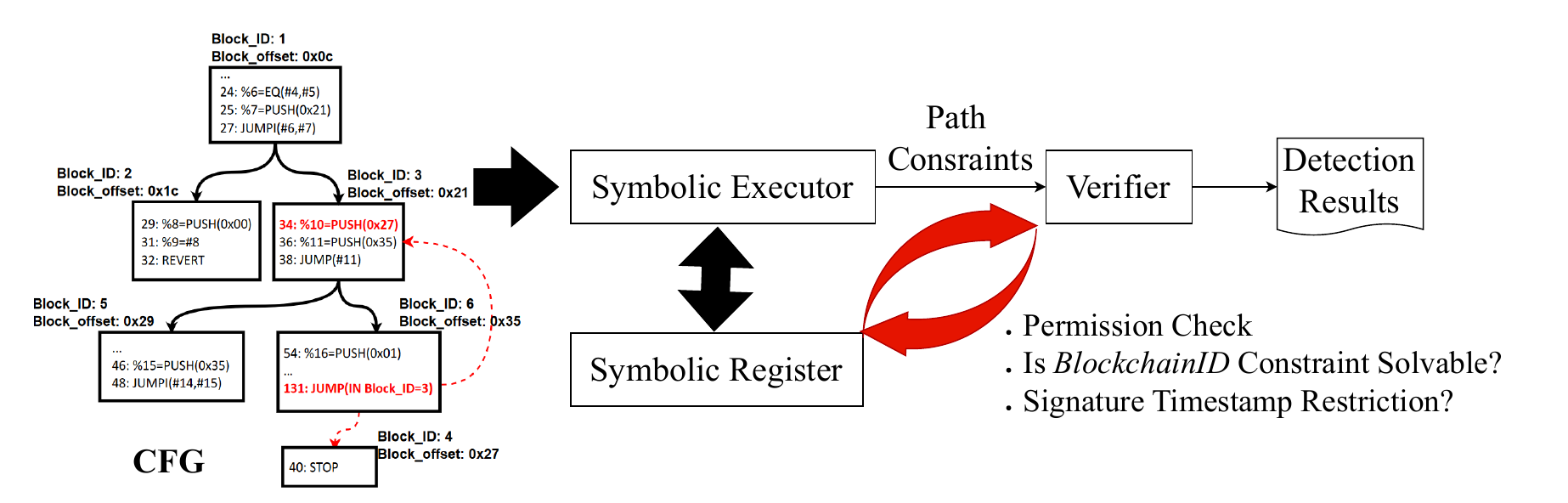}
    \caption{Symbolic Execution Verification for \textit{Cross-Chain Replay Attack} (CCRA).}
    \label{fig:example12}
\end{figure}
\vspace{-0.3cm}
 
 As shown in Figure~\ref{fig:example12} for the symbolic execution verification targeting \textit{Cross-Chain Replay Attack} (CCRA), the \textit{Symbolic Executor} traverses paths and collects path constraints. The \textit{Verifier} queries the \textit{Symbolic Register} to prove \textit{whether the \textit{BlockchainID} Constraint is solvable}, \textit{whether there are \textit{Signature Timestamp Restrictions}}, and \textit{whether there is permission verification of the caller's identity}. Solving these constraints ensures the reachability of paths and completes attribute verification. This symbolic modeling is key to a complete collection of path constraints, which can be used to check whether the contract address involved in the external contract function call is symbolic. Based on this, we determine whether the contract call may be externally controlled, effectively identifying potential function callback vulnerabilities. The completed path also supports cross-contract path constraint analysis, enhancing the security verification of contracts.

\subsection{\textit{EVM-Inequivalent Code Smells} Detection}
\label{subsec:detect_defects}
In this section, we detail the \textit{Domain-Specific Patterns} required for detection and how to guide the path search:

\textbf{(1) \textit{Cross-Chain Replay Attack} (CCRA).} \textit{EquivGuard} checks if the contract allows the logic of the function to be approved and executed by external parties via signatures, such as \textit{permit()} function. Furthermore, it examines whether the signature verification process includes \textit{Chain ID} and blockchain information validation. By locating calls to the \textit{ecrecover()} function within the contract, \textit{EquivGuard} can narrow down the scope of function search. Coupled with reverse taint analysis, it determines if the variable related to \textit{Chain ID} can be easily manipulated in the code. Symbolic execution is then used to verify path feasibility for accurate detection.

\textbf{(2) \textit{Time Discrepancy Trap} (TDT).} \textit{EquivGuard} analyzes whether the contract employs block generation numbers to calculate time consumption by locating code blocks involving \textit{block.number}. It performs reverse taint analysis to identify the sources and propagation paths of variables. \textit{EquivGuard} then determines if these variables are constants and examines the judgment conditions of the calculation results. Finally, symbolic execution verifies the feasibility of the identified path and potential inconsistencies in time calculation across different EVM-compatible blockchains, ensuring accurate detection.

\textbf{(3) \textit{Phishing Contract Address} (PCA).} \textit{EquivGuard} analyzes whether contracts contain unrestricted method calls to fixed contract addresses. To reduce false positives, \textit{EquivGuard} employs static taint analysis to track the propagation of these addresses within the contract. It uses symbolic execution to verify the feasibility of paths involving such calls. 

\textbf{(4) \textit{Gas Limit Imbalance} (GLI).} 
\textit{EquivGuard} analyzes contracts to detect operations setting specific gas limits. Specifically, there is a query for the remaining gas amount using the \textit{gasleft()} function~\cite{gasleft} and adjusts the logic accordingly. By employing I-PDG combined with static taint analysis, it determines whether such logic exists in the code and whether the restriction is subject to conditional checks. If such conditions are found, GLI is flagged.

\textbf{(5) \textit{Fixed Gas Reentrancy} (FGR).} \textit{EquivGuard} first analyzes whether the contract includes token transfers caused by the \textit{transfer()/send()} methods. Second, it examines whether the contract follows the safe development pattern of \textit{ Check->Effect->Interaction} (C-E-I)~\footnote{C-E-I requires ensuring timely state updates before interacting with external contracts, reducing the possibility of malicious contracts attempting to hijack control flow after external calls.}~\cite{cei,Unityis}.  If the contract includes \textit{transfer()/send()} methods but does not adhere to the safe \textit{Check->Effect->Interaction} development pattern, then \textit{EquivGuard} considers this smell to exist.

\textbf{(6) \textit{Block Height Misalignment} (BHM).} \textit{EquivGuard} analyzes whether the contract contains restrictions on specific block heights. It utilizes I-PDG to examine whether the contract's execution statements impose restrictions on specific block heights. Additionally, it performs taint analysis to track the propagation of variables related to block heights and determine their impact on control flow decisions. Furthermore, \textit{EquivGuard} employs symbolic execution to verify the feasibility of the identified paths that involve block height restrictions, ensuring accurate detection.

\section{Evaluation}
\label{sec:evaluation}
In this section, we analyze and evaluate the effectiveness of \textit{EquivGuard} in detecting \textit{EVM-Inequivalent Code Smells} by answering the following research questions:

\begin{itemize}
\setlength{\itemindent}{-6mm}
\item RQ1: How does \textit{EquivGuard} perform on real large-scale datasets?
\item RQ2: What is the performance of \textit{EquivGuard} in detecting \textit{EVM-Inequivalent Code Smells}?
\item RQ3: What is the impact of different phases within \textit{EquivGuard}?
\end{itemize}

{\bf Experimental Setup.}
The experiment was conducted on a computer running Ubuntu 20.04 LTS, equipped with a 16-core Intel(R) Xeon(R) Gold 5217 processor and 120 GB of memory. We use solc-select~\cite{solcselect} to switch compiler versions and query the asset through \textit{Etherscan}'s API~\cite{etherscanapi}.

\subsection{RQ1: Performance on large-scale datasets}
\label{sec:rq1}
\textbf{Dataset.} 
To analyze the performance of \textit{EquivGuard} on large-scale datasets and explore the distribution of \textit{EVM Inequivalent Code Smells} across different blockchain environments, we collected and experimented with 905,948 contract source codes from six mainstream EVM-compatible blockchains. These blockchains account for 80.64\% of the total crypto-market assets by the time of writing the paper~\cite{defilIama}, including \textit{Ethereum} (66.25\%), \textit{Binance Smart Chain} (BSC) (6.23\%), \textit{Arbitrum} (4.29\%), \textit{Polygon} (1.35\%), \textit{Optimism} (1.26\%), and \textit{Avalanche} (1.26\%). 

\textbf{Results.} Table~\ref{tab:RQ1} shows the statistical results of \textit{EquivGuard}’s detection on large-scale datasets across different blockchains. Overall, \textit{EVM-Inequivalent Code Smells} are widely distributed across various blockchains. 115,148 contracts on the \textit{Binance Smart Chain} (BSC) contain code smells, accounting for 31.80\%. \textit{Avalanche} follows with a proportion of 21.62\%, while \textit{Optimism} has the lowest proportion at 10.08\%. Among the various types of code smells, \textit{Phishing Contract Address} (PCA), \textit{Time Discrepancy Trap} (TDT), and \textit{Fixed Gas Reentrancy} (FGR) have the highest proportions, especially on the \textit{Binance Smart Chain} (BSC), where each of these three types accounts for nearly 10\%. \textit{Avalanche} shows a significant presence of \textit{Time Discrepancy Trap} (TDT) (2,333 cases, 6.37\%) and \textit{Fixed Gas Reentrancy} (FGR) (3,658 cases, 9.98\%). In contrast, \textit{Block Height Misalignment} (BHM) and \textit{Gas Limit Imbalance} (GLI) have proportions below 0.23\% across the six blockchains, indicating a general awareness among developers to avoid dependence on specific block height and block time. Overall, the average proportion of \textit{EVM Inequivalent Code Smells} across all blockchains is 17.70\%. These data suggest that \textit{EVM Inequivalent Code Smells} are prevalent in different blockchain environments, highlighting developers’ lack of awareness regarding these smells in reused contracts and the urgent need to strengthen preventive measures.

\begin{table}[htb]
    \centering
    \caption{Statistics of detection results of \textit{EquivGuard} in large-scale datasets}
    \resizebox{0.96\linewidth}{!}{%
\begin{tabular}{l|lllllll}
\hline
\textbf{Blockchain} & \textbf{CCRA} & \textbf{PCA}  & \textbf{BHM} & \textbf{TDT}   & \textbf{GLI} & \textbf{FGR}   & \textbf{Total}  \\ \hline
\textbf{Arbitrum}   & 771(1.26\%)   & 3152(5.16\%)  & 15(0.02\%)   & 1972(3.23\%)   & 32(0.05\%)   & 2363(3.87\%)   & 8305(13.59\%)   \\
\textbf{Avalanche}  & 443(1.21\%)   & 1441(3.93\%)  & 36(0.10\%)   & 2333(6.37\%)   & 10(0.03\%)   & 3658(9.98\%)   & 7921(21.62\%)   \\
\textbf{BSC}        & 1512(0.42\%)  & 35018(9.62\%) & 841(0.23\%)  & 39924(10.97\%) & 43(0.01\%)   & 37810(10.55\%) & 115148(31.80\%) \\
\textbf{Optimism}   & 210(1.63\%)   & 593(4.60\%)   & 3(0.02\%)    & 263(2.04\%)    & 13(0.10\%)   & 218(1.69\%)    & 1300(10.08\%)   \\
\textbf{Polygon}    & 2773(1.92\%)  & 7102(4.93\%)  & 183(0.13\%)  & 4219(2.93\%)   & 59(0.04\%)   & 6002(4.17\%)   & 20338(14.12\%)  \\
\textbf{Ethereum}   & 1073(0.37\%)  & 12706(4.42\%) & 138(0.05\%)  & 5277(1.84\%)   & 16(0.01\%)   & 23955(8.34\%)  & 43165(15.03\%)  \\ \hline
\end{tabular}
    }
    \label{tab:RQ1}
\end{table}

\textbf{Answer to RQ1:} Analysis of 905,948 contracts across six blockchains revealed the widespread presence of \textit{EVM Inequivalent Code Smells}, averaging 17.70\%. \textit{Phishing Contract Address} (PCA), \textit{Time Discrepancy Trap} (TDT), and \textit{Fixed Gas Reentrancy} (FGR) are particularly prominent on \textit{Binance Smart Chain} (BSC) and \textit{Avalanche}. The high prevalence of affected contracts underscores a major security threat and necessitates urgent developer risk prevention.

\subsection{RQ2: Evaluation of detection effects}

\textbf{Dataset.} To evaluate \textit{EquivGuard}'s performance in detecting \textit{EVM-Inequivalent Code Smells}, we randomly sampled and manually analyzed false positives and false negatives from the detection results to assess \textit{Precision} and \textit{Recall}. Following previous research~\cite{jiashuo}, we used a confidence interval-based sampling method with a 95\% confidence level and a 10-point confidence interval, resulting in 444 positives and 96 negatives from the detection results. Then, we manually checked these samples to determine the existence of smells, ultimately obtaining 426 true positive samples and 114 true negative samples.  Table~\ref{tab:RQ2} presents the number of samples for positive label contracts with different \textit{EVM-Inequivalent Code Smells}.

\vspace{-0.3cm}
\begin{table}[h]
    \centering
    \caption{Precision Analysis Results Statistics of \textit{EquivGuard}}
    \resizebox{0.7\linewidth}{!}{%
\begin{tabular}{l|llllll}
\hline
                   & \textbf{PCA} & \textbf{GLI} & \textbf{CCRA} & \textbf{FGR} & \textbf{TDT} & \textbf{BHM} \\ \hline
\textbf{Samples}   & 95           & 14           & 88            & 96           & 94           & 57           \\
\textbf{TP}        & 95           & 14           & 86            & 91           & 88           & 48           \\
\textbf{FP}        & 0            & 0            & 2             & 5            & 6            & 9            \\
\textbf{Precision} & 100.00\%     & 100.00\%     & 97.73\%       & 94.79\%      & 93.62\%      & 84.21\%      \\ \hline
\end{tabular}
    }
    \label{tab:RQ2}
\end{table}
\vspace{-0.3cm}

\subsubsection{\textbf{False Positive Analysis}} To analyze the precision of \textit{EquivGuard}, we randomly sampled the positive detection results according to different code smell categories and performed a false positive analysis. For each item, we calculate precision through the formula $Precision = TP/(TP+FP)$, and calculate the overall detection $ Precision_{(Overall)} =\frac{ \sum_{i =1}^{n}p_{c_{i}}\times |c_ {i}|
}{\sum_{i=1}^{n}|c_{i}|
} $, $p_{c_{i}}$ denotes the precision for code smell $i$, and $|c_i|$ represents the number of code smell $i$~\cite{yang2023definition}. 

\textbf{Results.}
According to the results in Table~\ref{tab:RQ2}, the $Precision_{(Overall)}$ of \textit{EquivGuard} reached 95.29\%, indicating that it has high accuracy in detecting \textit{EVM-Inequivalent Code Smells}. Through result analysis, we found that some false positives are related to the variability of permission control. Customized and diverse permission control is a key way to ensure contract security, but it also increases the complexity of analysis. For example, in the \textit{Block Height Misalignment} (BHM) detection, false positives mainly stem from identifying the key function calling permission. To reduce such problems as possible, we referred to the existing \textit{Path Protection Technologies} (PPTs)~\cite{ye2020clairvoyance}, but these technologies rely on static rule analysis, only cover the most common permission paths, and cannot guarantee the analysis of all situations. Furthermore, a small number of false positives are related to heuristic analysis invariants. To analyze contract invariants, we adopted a heuristic method of statically analyzing the read-and-write relationships of variables. However, this heuristic method will lead to inaccurate detection results and false positives when dealing with complex data structures such as dynamic arrays.

\subsubsection{\textbf{False Negative Analysis}} To evaluate the reliability of \textit{EquivGuard}'s detection, we calculate recall using $Recall = TP/(TP+FN)$ and perform cause analysis.

\textbf{Results.} Through manual inspection and analysis, the detection recall rate was calculated to be 99.06\%. We found 4 undetected (false negatives) code smells, all of which were \textit{Cross-Chain Replay Attack} (CCRA). These code smells mainly occur when calling the \textit{ecrecover()} for signature verification, which is implemented using assembly code. As EVM-compatible chains have limited support for opcodes and precompiled instructions, which restricts the scope for reusing contracts with assembly code. Research~\cite{assembly2022study} shows that manual inline assembly lacks generality, and the intricate logic raises developers' comprehension costs, making direct reuse less favorable. Furthermore, we plan to expand assembly code analysis capabilities in our subsequent work.

\textbf{Answer to RQ2:} \textit{EquivGuard} achieved 95.29\% Precision and 99.06\% Recall in \textit{EVM-Inequivalent Code Smells} detection, demonstrating high reliability. 

\subsection{RQ3: Ablation Experiment}

\textbf{Dataset and Group Settings.} To evaluate the necessity of Step 2's \textit{Domain-guided Reverse Path Search} (DRPS) and Step 3's \textit{Symbolic Execution Verification} (SEV) in \textit{EquivGuard}, we conducted a series of ablation experiments with four groups: \textbf{Neither} (both DRPS and SEV disabled), \textbf{No Path Guidance} (DRPS disabled), \textbf{No Symbolic Execution} (SEV disabled), and \textbf{Full Setup} (both DRPS and SEV enabled), with a unified time limit of 60 seconds. To ensure accuracy and reliability, we used the RQ2 dataset, which includes 426 positive samples and 114 negative samples, manually checked and analyzed to ensure the validity of the results.

\subsubsection{\textbf{Effectiveness Analysis.}} 
In the four experimental groups, the implementation of different functional components resulted in varying detection capabilities, each with distinct characteristics. The \textbf{Full Setup} group, where both DRPS and SEV were enabled, achieved the highest precision (95.05\%) and recall (99.06\%), indicating that the combination of domain-guided path search and symbolic execution path validation can achieve accurate and reliable detection. In contrast, disabling SEV in the \textbf{No Symbolic Execution} group resulted in a significant drop in recall (41.08\%) and precision (65.30\%), demonstrating the importance of symbolic execution in enhancing true positives and reducing false negatives. Similarly, disabling DRPS in the \textbf{No Path Guidance} group led to a substantial decline in recall (9.86\%) and precision (27.81\%), highlighting the crucial role of domain-guided path search in guiding analysis and reducing false positives. Finally, the \textbf{Neither} group, which disabled both DRPS and SEV, performed the worst, with precision (11.20\%) and recall (3.29\%) dropping to extremely low levels, further proving the necessity of domain-guided path search and symbolic execution path validation.

\begin{table}[htb]
    \centering
    \caption{Ablation Study Results}
    \label{tab:ablation_results}
    \resizebox{0.88\linewidth}{!}{%
    \begin{tabular}{l|llll}
    \hline
    \textbf{Metric} & \textbf{Full Setup} & \textbf{No Symbolic Execution} & \textbf{No Path Guidance} & \textbf{Neither} \\ \hline
    \# TP & 422 & 175 & 42 & 14 \\
    \# FP & 22 & 93 & 109 & 111 \\
    \# FN & 4 & 251 & 384 & 412 \\ 
    Precision (\%) & 95.05\% & 65.30\% & 27.81\% & 11.20\% \\
    Recall (\%) & 99.06\% & 41.08\% & 9.86\% & 3.29\% \\ 
    Average Time (s) & 45.2 s & 21.1 s & 57.2 s & 10.3 s \\ \hline
    \# Timeouts & 4 (0.74\%) & 0 (0) & 336 (62.22\%) & 0 (0) \\ \hline
    \end{tabular}
    }
\end{table}

\subsubsection{\textbf{Efficiency Analysis.}} We also analyzed the average time consumption under different experimental groups and found that \textit{EquivGuard}’s phased strategy helps combine the characteristics of different techniques to ensure efficient detection. The \textbf{Full Setup} group had an average time of 45.2 seconds, which is moderate. Although the \textbf{No Symbolic Execution} group had an average time of only 21.1 seconds, mainly due to reduced time for symbolic execution path traversal and constraint solving, this also resulted in a significant decline in detection effectiveness. The \textbf{No Path Guidance} group had the longest average time (57.2 seconds) and the highest number of timeouts (336), indicating that Step 2's \textit{Domain-guided Reverse Path Search} (DRPS) is crucial for alleviating the intensive computation and path explosion faced by symbolic execution path reachability verification. Although the \textbf{Neither} group had the shortest average time (10.3 seconds), its detection effectiveness was extremely poor.

\subsubsection{\textbf{Analysis of the Impact of Symbolic Execution Timeout.}}
Timeouts significantly impact the efficiency of symbolic execution, thus affecting the performance of \textit{EquivGuard}. Path explosion and blind path traversal can easily cause timeouts, limiting the scope of path exploration, reducing detection accuracy, and hindering scalability. As shown in Table~\ref{tab:ablation_results}, by comparing the \textbf{Full Setup} and \textbf{No Symbolic Execution} groups, it is evident that the introduction of symbolic execution significantly improves recall from 41.08\% to 99.06\%, and also enhances precision. Meanwhile, in the absence of path guidance during the symbolic execution's path reachability verification (\textbf{No Path Guidance}), detection precision and recall drop to 27.81\% and 9.86\%, respectively. Therefore, blind path traversal and path explosion can easily lead to timeouts. This not only limits the paths explored by symbolic execution but also may mistakenly identify unverified paths as issues due to interrupted verification, significantly increasing false positives. Additionally, the \textbf{No Path Guidance} group had the longest average analysis time (57.2 seconds) and the most timeouts (336), severely affecting efficiency. Frequent timeouts also expose scalability issues of \textit{EquivGuard} when handling large and complex programs, limiting its practical applicability. Our analysis of experimental results identified three main causes of timeouts: path explosion, constraint-solving complexity, and external dependencies. For a detailed analysis, please refer to Appendix A.

\textbf{Answer to RQ3:} The analysis shows that \textit{Domain-guided Reverse Path Search} (DRPS) aids in the traversal of paths by symbolic execution, while \textit{Symbolic Execution Verification} (SEV) effectively validates path reachability, improving both precision and recall. \textit{EquivGuard} combines the characteristics of different techniques to achieve efficient detection.

\section{Discussion}
\label{sec:discussion}

\subsection{Exploiting \textit{Cross-Chain Replay Attack} (CCRA) in \textit{Multichain} Project}
Through large-scale detection, \textit{EquivGuard} discovered that \textit{Multichain}'s cross-chain project is affected by \textit{Cross-Chain Replay Attack} (CCRA), compromising the security of its managed assets. In this subsection, we will analyze how the \textit{Cross-Chain Replay Attack} (CCRA) in the reused contract is exploited and the resulting asset risks. \textit{Multichain}'s contract contains the \textit{Cross-Chain Replay Attack} (CCRA), which arose from hard-coding the chainId as 122~\footnote{https://moonscan.io/address/0x818ec0a7fe18ff94269904fced6ae3dae6d6dc0b\#code\#L306}, while the actual deployed blockchain's chainId was 1284. This caused the failure of the verification mechanism intended to prevent replay attacks. This threat continuously affects assets on other chains where the same contracts are reused.

The overall exploitation process consists of three stages: (\uppercase\expandafter{\romannumeral1}) Retrieve signatures with \textit{ChainID} 122. Hackers can search for their previously used signature information from historical transactions of a blockchain with \textit{ChainID} 122 or create related signature information offline. (\uppercase\expandafter{\romannumeral2}) Verify signatures on the chain with \textit{ChainID} 1284. Hackers call the \textit{transferWithPermit()} function on the \textit{AnyswapV5ERC20} contract deployed on blockchain with \textit{ChainID} 1284 using the signatures. The core reason why the signature verification succeeds is due to a check using the incorrect \textit{ChainID} 122 in the signature verification logic. (\uppercase\expandafter{\romannumeral3}) Profit. Attackers transfer tokens to gain profit.

\begin{figure}[htb]
    \centering
    \includegraphics[width=0.90\columnwidth]{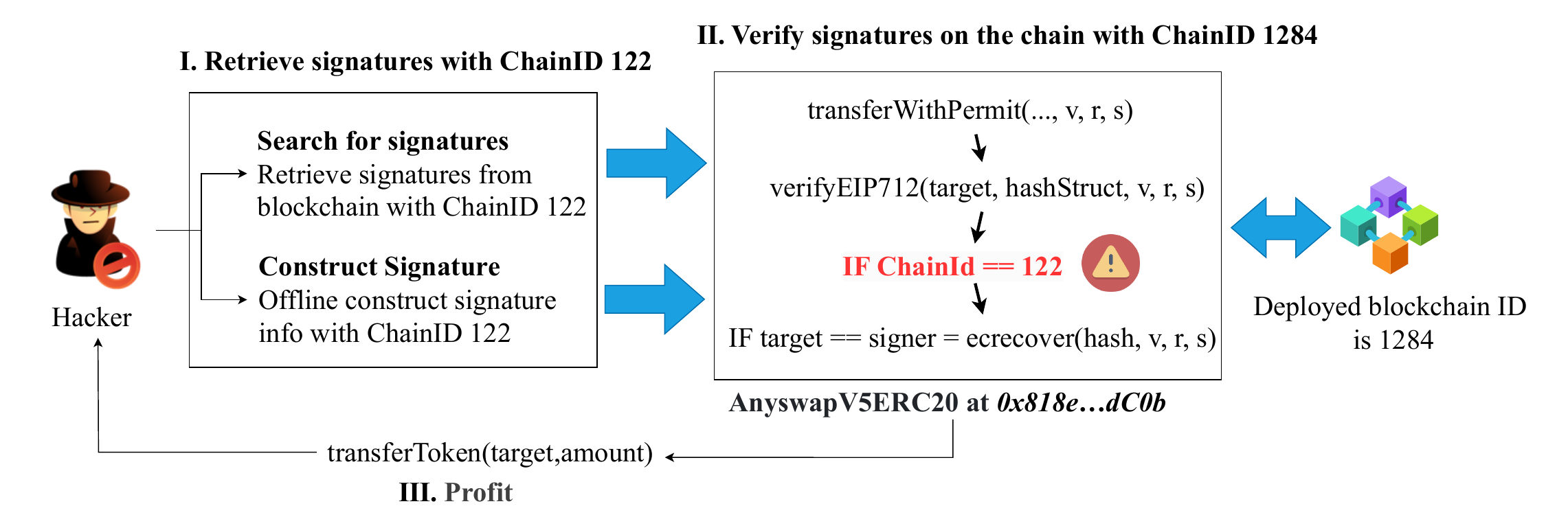}
    \caption{CCRA Exploitation in the \textit{Multichain}'s contract.}
    \label{fig:expolit}
\end{figure}

Attackers can search for exploitable historical transaction signatures and reuse them across different chains to generate fraudulent transactions. As \textit{Multichain}’s cross-chain router at \textit{0x818e...dC0b} used this contract code, approximately \$28.52 K in assets were put at risk. \textit{EquivGuard} promptly identified and reported this code smell.

\subsection{Implications}
\textit{EVM-inequivalent code smells} expose design flaws in contract reuse, leading to contract execution inconsistency. \textit{EquivGuard} automatically detects these code smells before reusing, improving the reliability and reusability of reused code. This study of \textit{EVM-inequivalent code smells} and \textit{EquivGuard} may provide new insights for secure contract reuse.

\textbf{\textit{For Auditors.}} Auditors can use \textit{EquivGuard} to mitigate risks from \textit{EVM-inequivalent code smells}. While code reuse enhances audit efficiency, it introduces new attack vectors when reused across blockchains. For instance, \textit{Wintermute}'s wallet attack~\cite{OP_replay} resulted in a \$20 million loss when Ethereum-safe code was reused on \textit{Optimism}. \textit{EquivGuard} helps auditors analyze multi-chain execution inconsistencies in reused contracts and provide comprehensive security guidance based on different blockchain designs (e.g., \textit{Gas Mechanisms} and \textit{Consensus Protocols}).

\textbf{\textit{For Developers.}} \textit{EquivGuard} helps developers avoid negative impacts due to the misunderstandings of \textit{directly copying smart contracts for multi-chain deployments}. While multi-chain reuse saves costs, it can introduce \textit{EVM-inequivalent code smells}, creating security threats. \textit{EquivGuard} alerts developers to these code smells, providing location and category information. This serves as a reminder for developers to conduct thorough testing in various execution environments to mitigate the risks associated with EVM-inequivalent execution.

\textbf{\textit{For Researchers.}} Researchers may further investigate programming practices that cause inconsistent execution in reused contracts to provide more comprehensive advice for development. For instance, they could develop a tool to automatically identify the negative impacts of \textit{EVM-inequivalent code smells}, offering users targeted improvement suggestions.

\textbf{\textit{For Community.}} The blockchain community can improve the execution layer of blockchains to mitigate risks. This could involve conducting a comprehensive analysis of the root causes of inconsistent execution stemming from reused contracts, considering different \textit{Gas Mechanisms} and \textit{Consensus Protocols}. Furthermore, it could involve implementing version updates to prevent the deployment of problematic contracts or to alert contract deployers.

\subsection{Threats to Validity}
\textbf{Internal Validity.} One internal threat is that not all code smells necessarily lead to financial loss or attacks; however, they indicate underlying design flaws that can create opportunities for vulnerabilities, especially in new scenarios from multi-chain contract reuse. Developers urgently need to enhance their awareness of \textit{EVM-Inequivalent Code Smells}, especially when reusing contracts across multiple chains. The other internal threat is using \textit{Domain-Specific Patterns} to guide path searches. These patterns, derived from security development experience, are crucial for taint propagation analysis. For instance, in \textit{Cross-Chain Replay Attack} (CCRA) detection, \textit{ecrecover()} calls are identified as sinks, as they are essential for signature verification. The analysis traces taint flow from external inputs (sources) to these sinks, incorporating \textit{chain ID} checks. They provide necessary instructions to help identify sources and sinks and indicate code smell presence. 

\noindent \textbf{External Validity.} One external threat is the generalizability of the code smell definition. To ensure the identified smells reflect real developer issues, we analyzed 1,379 security audit reports and 326 \textit{Stack Overflow} posts related to contract reuse. This work is the first to define and detect \textit{EVM-Inequivalent Code Smells}, tracing the identified smells back to their sources in \textit{Stack Overflow} and security audits. Another external threat is the risk of incomplete coverage, as using a single tag may result in missing relevant posts or including irrelevant ones. To ensure data completeness, we used an extended combination of related tags, including ``\textit{EVM}", ``\textit{EVM Equivalent}", ``\textit{EVM Compatible}". This broader tag selection helped capture more relevant posts and minimize omissions. Two rounds of manual reviews were conducted to eliminate irrelevant content.

\vspace{-0.15cm}
\section{Related Work}
\label{sec:rw}
\subsection{Defining and Detecting Bugs in Smart Contracts}

Based on both real-world attacks and theoretical research, researchers have proposed many solutions to address the proliferation of bugs. For example, Luu et al.~\cite{makingsmart} combined four vulnerability patterns and proposed the symbolic execution detection tool \textit{Oyente} based on bytecode analysis, providing a classic solution for contract vulnerability detection. By analyzing posts on \textit{Ethereum StackExchange}~\cite{stackex_eth} and real-world smart contracts, Chen et al. summarized 20 types of smart contract defects and highlighted 5 high-risk defects caused by protocol errors, providing developers with identification and guidance on fixing contract deficiencies~\cite{chen2020defining}. Zhang et al.~\cite{jiashuo} through large-scale analysis of \textit{Ethereum} transactions, smart contracts, and \textit{StackExchange} posts, found that developers face 5 main types of obstacles when dealing with crypto-related tasks and surveyed industry insiders to reveal the root causes of these obstacles and suggest improvements, providing practical guidance to improve the encryption task development experience. Yang et al.~\cite{yang2023definition} defined and explained 5 common defect types in \textit{Non-Fungible Token} (NFT) contracts by analyzing \textit{StackOverflow} posts and proposed a symbolic execution-based tool, \textit{NFTGuard}, to automatically detect these using contract \textit{Abstract Syntax Tree} (AST) and bytecode features.

\subsection{Research on Smart Contract Reuse}
Chen et al.~\cite{chenxiang_reuse} analyzed 146,452 open-source \textit{Ethereum} contracts, finding widespread code reuse with ERC20 tokens being the most commonly reused. Sun et al.~\cite{sun2023demystifying} studied over 350,000 contracts, observing that 50\% of self-developed sub-contracts had duplicate functions, while 35\% of external sub-contracts had issues such as inconsistent usage. They also extracted 61 frequent reuse patterns to guide secure contract development.
Pierro et al.~\cite{reuse1} compared mainstream smart contract corpora, noting that only a small portion reused code from the secure \textit{OpenZeppelin} repository~\cite{openzepplin}. They recommended leveraging the \textit{OpenZeppelin} Solidity Library to improve contract security. Huang et al.~\cite{reuseckg} constructed a semantic\textit{ Code Knowledge Graph} to uncover unknown factors in smart contract reuse, effectively enhancing code recommendation accuracy and developer efficiency.

\textbf{\textit{Differences.}} First, this is the first study to focus on the inconsistent execution of reused contracts across EVM-compatible blockchains. While reusing \textit{Ethereum} contracts on different platforms introduces new attack vectors, research on \textit{EVM-Inequivalent Code Smells} has been lacking. Secondly, our analysis combines insights from \textit{StackOverflow} posts and security audit reports from security companies, enabling us to better collect and analyze real-world cases. Thirdly, to detect \textit{EVM-Inequivalent Code Smells}, we propose a new method that combines dynamic and static techniques. This approach aims to achieve more efficient detection by leveraging the efficient path search of static taint analysis and the verification capabilities of symbolic execution, thereby improving the comprehensiveness and accuracy of the detection process.

\vspace{-0.15cm}
\section{Conclusion}
\label{sec:conclusion}
In this paper, we conduct the analysis of \textit{EVM-Inequivalent Code Smells}, which is the first study on inconsistent execution of \textit{Ethereum} contracts when reused on EVM-compatible blockchains. By analyzing 1,379 security audit reports and 326 \textit{Stack Overflow} posts, we identified and defined six \textit{EVM-Inequivalent Code Smells}. Discovering these code smells, which cause inconsistent execution of reused contracts on multi-chains, ensures more secure and reliable contract reuse. To aid in discovering reused contracts that contain \textit{EVM-Inequivalent Code Smells} in the real world, we designed \textit{EquivGuard} to implement automated detection of contracts. \textit{EquivGuard} identifies key paths by leveraging \textit{Domain-Specific Patterns}, which encapsulate domain knowledge for analyzing various code smells, and combines symbolic execution to verify path reachability. We conducted large-scale experiments on 905,948 contracts across six mainstream blockchains, achieving a detection precision of 95.29\%. Additionally, our experiments revealed that 17.70\% of contracts contained at least one such code smell. This highlights the prevalence of these code smells and reminds developers to prevent inconsistencies when reusing smart contracts.

\vspace{-0.15cm}
\section{Data Availability}
We have anonymized and made \textit{EquivGuard}'s source code, empirical study materials, experimental datasets, and results publicly available at \href{https://anonymous.4open.science/r/EquivGuard-68B0}{https://anonymous.4open.science/r/EquivGuard-68B0}.

\begin{acks}
    This research is supported by the National Key Research and Development Program of China (2022YFB2703204), the National Natural Science Foundation of China (No. 62032025, No. 62302534), the Guangdong Basic and Applied Basic Research Foundation (No. 2025A1515011632), and the Major Key Project of Peng Cheng Laboratory under Grant PCL2023A05-2.
\end{acks}

\bibliographystyle{ACM-Reference-Format}
\bibliography{ref}


\begin{thebibliography}{56}


\ifx \showCODEN    \undefined \def \showCODEN     #1{\unskip}     \fi
\ifx \showDOI      \undefined \def \showDOI       #1{#1}\fi
\ifx \showISBNx    \undefined \def \showISBNx     #1{\unskip}     \fi
\ifx \showISBNxiii \undefined \def \showISBNxiii  #1{\unskip}     \fi
\ifx \showISSN     \undefined \def \showISSN      #1{\unskip}     \fi
\ifx \showLCCN     \undefined \def \showLCCN      #1{\unskip}     \fi
\ifx \shownote     \undefined \def \shownote      #1{#1}          \fi
\ifx \showarticletitle \undefined \def \showarticletitle #1{#1}   \fi
\ifx \showURL      \undefined \def \showURL       {\relax}        \fi
\providecommand\bibfield[2]{#2}
\providecommand\bibinfo[2]{#2}
\providecommand\natexlab[1]{#1}
\providecommand\showeprint[2][]{arXiv:#2}

\bibitem[0x52(2024)]%
        {GasLimit}
\bibfield{author}{\bibinfo{person}{0x52}.} \bibinfo{year}{2024}\natexlab{}.
\newblock \bibinfo{title}{An attacker can lock operator out of the pod by setting gas limit that's higher than the block gas limit of dest chain.}
\newblock \bibinfo{howpublished}{\url{https://solodit.xyz/issues/h-01-an-attacker-can-lock-operator-out-of-the-pod-by-setting-gas-limit-thats-higher-than-the-block-gas-limit-of-dest-chain-code4rena-holograph-holograph-contest-git}}.
\newblock


\bibitem[Academy(2023)]%
        {gasleft}
\bibfield{author}{\bibinfo{person}{Solidity Academy}.} \bibinfo{year}{2023}\natexlab{}.
\newblock \bibinfo{title}{Understanding Ethereum Gas: A Technical Deep Dive into gasleft()}.
\newblock \bibinfo{howpublished}{\url{https://medium.com/@solidity101/understanding-ethereum-gas-a-technical-deep-dive-into-gasleft-b0842742fd12}}.
\newblock


\bibitem[{Alex Beregszaszi (@axic), Jacques Wagener (@jacqueswww)}(2018)]%
        {eip1380}
\bibfield{author}{\bibinfo{person}{{Alex Beregszaszi (@axic), Jacques Wagener (@jacqueswww)}}.} \bibinfo{year}{2018}\natexlab{}.
\newblock \bibinfo{title}{EIP-1380: Reduced gas cost for call to self}.
\newblock \bibinfo{howpublished}{\url{https://eips.ethereum.org/EIPS/eip-1380}}.
\newblock


\bibitem[{Arbitrum}(2024)]%
        {arbitrum}
\bibfield{author}{\bibinfo{person}{{Arbitrum}}.} \bibinfo{year}{2024}\natexlab{}.
\newblock \bibinfo{title}{Arbitrum — The Future of Ethereum}.
\newblock \bibinfo{howpublished}{\url{https://arbitrum.io/}}.
\newblock


\bibitem[AuditBase(2024)]%
        {TDTcase}
\bibfield{author}{\bibinfo{person}{AuditBase}.} \bibinfo{year}{2024}\natexlab{}.
\newblock \bibinfo{title}{Consider using block.number instead of block.timestamp.}
\newblock \bibinfo{howpublished}{\url{https://detectors.auditbase.com/blocknumber-vs-timestamp-solidity}}.
\newblock


\bibitem[AuditOne(2024)]%
        {Blockheigth}
\bibfield{author}{\bibinfo{person}{AuditOne}.} \bibinfo{year}{2024}\natexlab{}.
\newblock \bibinfo{title}{Lack of Validation for valid\_till\_block\_height on FastBridge Service.}
\newblock \bibinfo{howpublished}{\url{https://solodit.xyz/issues/lack-of-validation-for-valid_till_block_height-on-fastbridge-service-auditone-none-aurorafastbridge-markdown}}.
\newblock


\bibitem[{BlockSec}(2024)]%
        {blocksec}
\bibfield{author}{\bibinfo{person}{{BlockSec}}.} \bibinfo{year}{2024}\natexlab{}.
\newblock \bibinfo{title}{Ensuring a Secure and Seamless Web3 World}.
\newblock \bibinfo{howpublished}{\url{https://blocksec.com/}}.
\newblock


\bibitem[{BNB Chain community}(2024)]%
        {bep}
\bibfield{author}{\bibinfo{person}{{BNB Chain community}}.} \bibinfo{year}{2024}\natexlab{}.
\newblock \bibinfo{title}{About the BEP Category}.
\newblock \bibinfo{howpublished}{\url{https://forum.bnbchain.org/t/about-the-bep-category/624}}.
\newblock


\bibitem[{CaptPython}(2019)]%
        {cei}
\bibfield{author}{\bibinfo{person}{{CaptPython}}.} \bibinfo{year}{2019}\natexlab{}.
\newblock \bibinfo{title}{Design pattern Checks-Effects-Interactions Pattern}.
\newblock \bibinfo{howpublished}{\url{https://ethereum.stackexchange.com/questions/66456/design-pattern-checks-effects-interactions-pattern}}.
\newblock


\bibitem[{CFI Team}(2024)]%
        {hardfork}
\bibfield{author}{\bibinfo{person}{{CFI Team}}.} \bibinfo{year}{2024}\natexlab{}.
\newblock \bibinfo{title}{Hard Forks}.
\newblock \bibinfo{howpublished}{\url{https://corporatefinanceinstitute.com/resources/cryptocurrency/hard-fork/}}.
\newblock


\bibitem[Chaliasos et~al\mbox{.}(2022)]%
        {assembly2022study}
\bibfield{author}{\bibinfo{person}{Stefanos Chaliasos}, \bibinfo{person}{Arthur Gervais}, {and} \bibinfo{person}{Benjamin Livshits}.} \bibinfo{year}{2022}\natexlab{}.
\newblock \showarticletitle{A study of inline assembly in solidity smart contracts}.
\newblock \bibinfo{journal}{\emph{Proc. ACM Program. Lang.}} \bibinfo{volume}{6}, \bibinfo{number}{OOPSLA2}, Article \bibinfo{articleno}{165} (\bibinfo{date}{Oct.} \bibinfo{year}{2022}), \bibinfo{numpages}{27}~pages.
\newblock
\urldef\tempurl%
\url{https://doi.org/10.1145/3563328}
\showDOI{\tempurl}


\bibitem[{Che Kohler}(2022)]%
        {BCT}
\bibfield{author}{\bibinfo{person}{{Che Kohler}}.} \bibinfo{year}{2022}\natexlab{}.
\newblock \bibinfo{title}{What Is Bitcoin Block Time?}
\newblock \bibinfo{howpublished}{\url{https://thebitcoinmanual.com/articles/btc-block-time/}}.
\newblock


\bibitem[Chen et~al\mbox{.}(2022)]%
        {chen2020defining}
\bibfield{author}{\bibinfo{person}{Jiachi Chen}, \bibinfo{person}{Xin Xia}, \bibinfo{person}{David Lo}, \bibinfo{person}{John Grundy}, \bibinfo{person}{Xiapu Luo}, {and} \bibinfo{person}{Ting Chen}.} \bibinfo{year}{2022}\natexlab{}.
\newblock \showarticletitle{Defining Smart Contract Defects on Ethereum}.
\newblock \bibinfo{journal}{\emph{IEEE Transactions on Software Engineering}} \bibinfo{volume}{48}, \bibinfo{number}{1} (\bibinfo{year}{2022}), \bibinfo{pages}{327--345}.
\newblock
\urldef\tempurl%
\url{https://doi.org/10.1109/TSE.2020.2989002}
\showDOI{\tempurl}


\bibitem[Chen et~al\mbox{.}(2021)]%
        {chenxiang_reuse}
\bibfield{author}{\bibinfo{person}{Xiangping Chen}, \bibinfo{person}{Peiyong Liao}, \bibinfo{person}{Yixin Zhang}, \bibinfo{person}{Yuan Huang}, {and} \bibinfo{person}{Zibin Zheng}.} \bibinfo{year}{2021}\natexlab{}.
\newblock \showarticletitle{Understanding Code Reuse in Smart Contracts}. In \bibinfo{booktitle}{\emph{2021 IEEE International Conference on Software Analysis, Evolution and Reengineering (SANER)}}. \bibinfo{pages}{470--479}.
\newblock
\urldef\tempurl%
\url{https://doi.org/10.1109/SANER50967.2021.00050}
\showDOI{\tempurl}


\bibitem[{Consensys}(2019)]%
        {transfersend}
\bibfield{author}{\bibinfo{person}{{Consensys}}.} \bibinfo{year}{2019}\natexlab{}.
\newblock \bibinfo{title}{Stop Using Solidity's transfer() Now}.
\newblock \bibinfo{howpublished}{\url{https://consensys.io/diligence/blog/2019/09/stop-using-soliditys-transfer-now/}}.
\newblock


\bibitem[{ConsenSys}(2024)]%
        {consensys}
\bibfield{author}{\bibinfo{person}{{ConsenSys}}.} \bibinfo{year}{2024}\natexlab{}.
\newblock \bibinfo{title}{Consensys - A complete suite of trusted products to build anything in web3}.
\newblock \bibinfo{howpublished}{\url{https://consensys.io/}}.
\newblock


\bibitem[Crytic(2024)]%
        {solcselect}
\bibfield{author}{\bibinfo{person}{Crytic}.} \bibinfo{year}{2024}\natexlab{}.
\newblock \bibinfo{title}{Manage and switch between Solidity compiler versions}.
\newblock \bibinfo{howpublished}{\url{https://github.com/crytic/solc-select}}.
\newblock


\bibitem[Cyfrin(2025)]%
        {solodit}
\bibfield{author}{\bibinfo{person}{Cyfrin}.} \bibinfo{year}{2025}\natexlab{}.
\newblock \bibinfo{title}{Solodit}.
\newblock \bibinfo{howpublished}{\url{https://solodit.cyfrin.io/}}.
\newblock


\bibitem[DefiLIama(2024)]%
        {defilIama}
\bibfield{author}{\bibinfo{person}{DefiLIama}.} \bibinfo{year}{2024}\natexlab{}.
\newblock \bibinfo{title}{DefiLiama EVM Chains}.
\newblock \bibinfo{howpublished}{Retrieved from \url{https://defillama.com/chains/EVM}}.
\newblock


\bibitem[Etherscan(2024)]%
        {etherscanapi}
\bibfield{author}{\bibinfo{person}{Etherscan}.} \bibinfo{year}{2024}\natexlab{}.
\newblock \bibinfo{title}{Etherscan APIs- Ethereum (ETH) API Provider}.
\newblock \bibinfo{howpublished}{\url{https://etherscan.io/apis}}.
\newblock


\bibitem[Etherscan(2025)]%
        {etherscanaudit}
\bibfield{author}{\bibinfo{person}{Etherscan}.} \bibinfo{year}{2025}\natexlab{}.
\newblock \bibinfo{title}{Etherscan Smart Contracts Audit and Security}.
\newblock \bibinfo{howpublished}{\url{https://etherscan.io/directory/Smart_Contracts/Smart_Contracts_Audit_And_Security}}.
\newblock


\bibitem[{evm.storage}(2024)]%
        {evmcode}
\bibfield{author}{\bibinfo{person}{{evm.storage}}.} \bibinfo{year}{2024}\natexlab{}.
\newblock \bibinfo{title}{An Ethereum Virtual Machine Opcodes Interactive Reference}.
\newblock \bibinfo{howpublished}{\url{https://www.evm.codes/}}.
\newblock


\bibitem[Exchange(2024)]%
        {stackex_eth}
\bibfield{author}{\bibinfo{person}{Ethereum~Stack Exchange}.} \bibinfo{year}{2024}\natexlab{}.
\newblock \bibinfo{title}{Ethereum Stack Exchange}.
\newblock \bibinfo{howpublished}{\url{https://ethereum.stackexchange.com/}}.
\newblock


\bibitem[{Gopal Gurram}(2021)]%
        {SOpost}
\bibfield{author}{\bibinfo{person}{{Gopal Gurram}}.} \bibinfo{year}{2021}\natexlab{}.
\newblock \bibinfo{title}{Can we deploy same ERC20-token on different blockchains?}
\newblock \bibinfo{howpublished}{\url{https://stackoverflow.com/questions/68802705/can-we-deploy-same-erc20-token-on-different-blockchains/68805158\#68805158}}.
\newblock


\bibitem[Huang et~al\mbox{.}(2023)]%
        {reuseckg}
\bibfield{author}{\bibinfo{person}{Qing Huang}, \bibinfo{person}{Dianshu Liao}, \bibinfo{person}{Zhenchang Xing}, \bibinfo{person}{Zhengkang Zuo}, \bibinfo{person}{Changjing Wang}, {and} \bibinfo{person}{Xin Xia}.} \bibinfo{year}{2023}\natexlab{}.
\newblock \showarticletitle{Semantic-Enriched Code Knowledge Graph to Reveal Unknowns in Smart Contract Code Reuse}.
\newblock \bibinfo{journal}{\emph{ACM Trans. Softw. Eng. Methodol.}} \bibinfo{volume}{32}, \bibinfo{number}{6}, Article \bibinfo{articleno}{147} (\bibinfo{date}{Sept.} \bibinfo{year}{2023}), \bibinfo{numpages}{37}~pages.
\newblock
\showISSN{1049-331X}
\urldef\tempurl%
\url{https://doi.org/10.1145/3597206}
\showDOI{\tempurl}


\bibitem[{Inspex}(2024)]%
        {OP_replay}
\bibfield{author}{\bibinfo{person}{{Inspex}}.} \bibinfo{year}{2024}\natexlab{}.
\newblock \bibinfo{title}{How 20 Million \$OP Was Stolen from the Multisig Wallet (Not Yet) Owned by Wintermute}.
\newblock \bibinfo{howpublished}{\url{https://inspexco.medium.com/how-20-million-op-was-stolen-from-the-multisig-wallet-not-yet-owned-by-wintermute-3f6c75db740a}}.
\newblock


\bibitem[Jia and Yin(2022)]%
        {jia2022evm_compat}
\bibfield{author}{\bibinfo{person}{Ruizhe Jia} {and} \bibinfo{person}{Steven Yin}.} \bibinfo{year}{2022}\natexlab{}.
\newblock \showarticletitle{To EVM or Not to EVM: Blockchain Compatibility and Network Effects}. In \bibinfo{booktitle}{\emph{Proceedings of the 2022 ACM CCS Workshop on Decentralized Finance and Security}} (Los Angeles, CA, USA) \emph{(\bibinfo{series}{DeFi'22})}. \bibinfo{publisher}{Association for Computing Machinery}, \bibinfo{address}{New York, NY, USA}, \bibinfo{pages}{23–29}.
\newblock
\showISBNx{9781450398824}
\urldef\tempurl%
\url{https://doi.org/10.1145/3560832.3563442}
\showDOI{\tempurl}


\bibitem[Luu et~al\mbox{.}(2016)]%
        {makingsmart}
\bibfield{author}{\bibinfo{person}{Loi Luu}, \bibinfo{person}{Duc-Hiep Chu}, \bibinfo{person}{Hrishi Olickel}, \bibinfo{person}{Prateek Saxena}, {and} \bibinfo{person}{Aquinas Hobor}.} \bibinfo{year}{2016}\natexlab{}.
\newblock \showarticletitle{Making Smart Contracts Smarter}. In \bibinfo{booktitle}{\emph{Proceedings of the 2016 ACM SIGSAC Conference on Computer and Communications Security}} (Vienna, Austria) \emph{(\bibinfo{series}{CCS '16})}. \bibinfo{publisher}{Association for Computing Machinery}, \bibinfo{address}{New York, NY, USA}, \bibinfo{pages}{254–269}.
\newblock
\showISBNx{9781450341394}
\urldef\tempurl%
\url{https://doi.org/10.1145/2976749.2978309}
\showDOI{\tempurl}


\bibitem[Ma et~al\mbox{.}(2022)]%
        {ma2021pluto}
\bibfield{author}{\bibinfo{person}{Fuchen Ma}, \bibinfo{person}{Zhenyang Xu}, \bibinfo{person}{Meng Ren}, \bibinfo{person}{Zijing Yin}, \bibinfo{person}{Yuanliang Chen}, \bibinfo{person}{Lei Qiao}, \bibinfo{person}{Bin Gu}, \bibinfo{person}{Huizhong Li}, \bibinfo{person}{Yu Jiang}, {and} \bibinfo{person}{Jiaguang Sun}.} \bibinfo{year}{2022}\natexlab{}.
\newblock \showarticletitle{Pluto: Exposing Vulnerabilities in Inter-Contract Scenarios}.
\newblock \bibinfo{journal}{\emph{IEEE Transactions on Software Engineering}} \bibinfo{volume}{48}, \bibinfo{number}{11} (\bibinfo{year}{2022}), \bibinfo{pages}{4380--4396}.
\newblock
\urldef\tempurl%
\url{https://doi.org/10.1109/TSE.2021.3117966}
\showDOI{\tempurl}


\bibitem[{Martin Lundfall (@Mrchico)}(2020)]%
        {eip2612}
\bibfield{author}{\bibinfo{person}{{Martin Lundfall (@Mrchico)}}.} \bibinfo{year}{2020}\natexlab{}.
\newblock \bibinfo{title}{ERC-2612: Permit Extension for EIP-20 Signed Approvals}.
\newblock \bibinfo{howpublished}{\url{https://eips.ethereum.org/EIPS/eip-2612}}.
\newblock


\bibitem[Nawaz(2012)]%
        {cardsorting1}
\bibfield{author}{\bibinfo{person}{Ather Nawaz}.} \bibinfo{year}{2012}\natexlab{}.
\newblock \showarticletitle{A Comparison of Card-sorting Analysis Methods}. In \bibinfo{booktitle}{\emph{APCHI '12. Proceedings of the 10th Asia Pacific Conference on Computer-Human Interaction}}, Vol.~\bibinfo{volume}{2}. \bibinfo{publisher}{Association for Computing Machinery}, \bibinfo{address}{United States}, \bibinfo{pages}{583--592}.
\newblock
\showISBNx{9784990656201}
\urldef\tempurl%
\url{http://apchi2012.org/}
\showURL{%
\tempurl}
\newblock
\shownote{The 10th Asia Pacific Conference on Computer Human Interaction. 2012 ; Conference date: 28-08-2012 Through 31-08-2012}.


\bibitem[OpenZeppelin(2024)]%
        {openzepplin}
\bibfield{author}{\bibinfo{person}{OpenZeppelin}.} \bibinfo{year}{2024}\natexlab{}.
\newblock \bibinfo{title}{openzeppelin-contracts}.
\newblock \bibinfo{howpublished}{\url{https://github.com/OpenZeppelin/openzeppelin-contracts}}.
\newblock


\bibitem[{Orderly Network}(2023)]%
        {evm_compatible}
\bibfield{author}{\bibinfo{person}{{Orderly Network}}.} \bibinfo{year}{2023}\natexlab{}.
\newblock \bibinfo{title}{What is an EVM Compatible Chain?}
\newblock \bibinfo{howpublished}{\url{https://medium.com/@orderlynetwork/what-is-an-evm-compatible-chain-46a7825adc4d}}.
\newblock


\bibitem[Pierro and Tonelli(2021)]%
        {reuse1}
\bibfield{author}{\bibinfo{person}{Giuseppe~Antonio Pierro} {and} \bibinfo{person}{Roberto Tonelli}.} \bibinfo{year}{2021}\natexlab{}.
\newblock \showarticletitle{Analysis of Source Code Duplication in Ethreum Smart Contracts}. In \bibinfo{booktitle}{\emph{2021 IEEE International Conference on Software Analysis, Evolution and Reengineering (SANER)}}. \bibinfo{pages}{701--707}.
\newblock
\urldef\tempurl%
\url{https://doi.org/10.1109/SANER50967.2021.00089}
\showDOI{\tempurl}


\bibitem[{SlowMist}(2024)]%
        {slowmist}
\bibfield{author}{\bibinfo{person}{{SlowMist}}.} \bibinfo{year}{2024}\natexlab{}.
\newblock \bibinfo{title}{SlowMist - Focusing on Blockchain Ecosystem Security}.
\newblock \bibinfo{howpublished}{\url{https://www.slowmist.com/}}.
\newblock


\bibitem[Soliditydeveloper(2022)]%
        {ecrecover}
\bibfield{author}{\bibinfo{person}{Soliditydeveloper}.} \bibinfo{year}{2022}\natexlab{}.
\newblock \bibinfo{title}{What is ecrecover in Solidity?}
\newblock \bibinfo{howpublished}{\url{https://soliditydeveloper.com/ecrecover}}.
\newblock


\bibitem[{Stackoveflow}(2024)]%
        {stackoveflow}
\bibfield{author}{\bibinfo{person}{{Stackoveflow}}.} \bibinfo{year}{2024}\natexlab{}.
\newblock \bibinfo{title}{Stack Overflow - Where Developers Learn, Share, \& Build Careers}.
\newblock \bibinfo{howpublished}{\url{https://stackoverflow.com/}}.
\newblock


\bibitem[Sun et~al\mbox{.}(2023)]%
        {sun2023demystifying}
\bibfield{author}{\bibinfo{person}{Kairan Sun}, \bibinfo{person}{Zhengzi Xu}, \bibinfo{person}{Chengwei Liu}, \bibinfo{person}{Kaixuan Li}, {and} \bibinfo{person}{Yang Liu}.} \bibinfo{year}{2023}\natexlab{}.
\newblock \showarticletitle{Demystifying the Composition and Code Reuse in Solidity Smart Contracts}. In \bibinfo{booktitle}{\emph{Proceedings of the 31st ACM Joint European Software Engineering Conference and Symposium on the Foundations of Software Engineering}} (San Francisco, CA, USA) \emph{(\bibinfo{series}{ESEC/FSE 2023})}. \bibinfo{publisher}{Association for Computing Machinery}, \bibinfo{address}{New York, NY, USA}, \bibinfo{pages}{796–807}.
\newblock
\showISBNx{9798400703270}
\urldef\tempurl%
\url{https://doi.org/10.1145/3611643.3616270}
\showDOI{\tempurl}


\bibitem[{Tanish Gupta}(2023)]%
        {2accounts}
\bibfield{author}{\bibinfo{person}{{Tanish Gupta}}.} \bibinfo{year}{2023}\natexlab{}.
\newblock \bibinfo{title}{EOAs vs Contracts: Understanding the Two Types of Ethereum Accounts}.
\newblock \bibinfo{howpublished}{\url{https://medium.com/\@tanish_gupta/eoas-vs-contracts-understanding-the-two-types-of-ethereum-accounts-378f9402d0e8}}.
\newblock


\bibitem[{Trail of Bits}(2024)]%
        {TrailofBits}
\bibfield{author}{\bibinfo{person}{{Trail of Bits}}.} \bibinfo{year}{2024}\natexlab{}.
\newblock \bibinfo{title}{Trail of Bits}.
\newblock \bibinfo{howpublished}{\url{https://www.trailofbits.com/}}.
\newblock


\bibitem[{TRON}(2024)]%
        {tron}
\bibfield{author}{\bibinfo{person}{{TRON}}.} \bibinfo{year}{2024}\natexlab{}.
\newblock \bibinfo{title}{Decentralize The Web}.
\newblock \bibinfo{howpublished}{\url{https://tron.network/}}.
\newblock


\bibitem[Tufano et~al\mbox{.}(2015)]%
        {codesmell}
\bibfield{author}{\bibinfo{person}{Michele Tufano}, \bibinfo{person}{Fabio Palomba}, \bibinfo{person}{Gabriele Bavota}, \bibinfo{person}{Rocco Oliveto}, \bibinfo{person}{Massimiliano Di~Penta}, \bibinfo{person}{Andrea De~Lucia}, {and} \bibinfo{person}{Denys Poshyvanyk}.} \bibinfo{year}{2015}\natexlab{}.
\newblock \showarticletitle{When and Why Your Code Starts to Smell Bad}. In \bibinfo{booktitle}{\emph{2015 IEEE/ACM 37th IEEE International Conference on Software Engineering}}, Vol.~\bibinfo{volume}{1}. \bibinfo{pages}{403--414}.
\newblock
\urldef\tempurl%
\url{https://doi.org/10.1109/ICSE.2015.59}
\showDOI{\tempurl}


\bibitem[Uniswap(2021)]%
        {uniswaprouter}
\bibfield{author}{\bibinfo{person}{Uniswap}.} \bibinfo{year}{2021}\natexlab{}.
\newblock \bibinfo{title}{Router02}.
\newblock \bibinfo{howpublished}{\url{https://docs.uniswap.org/contracts/v2/reference/smart-contracts/router-02}}.
\newblock


\bibitem[Vechain(2023)]%
        {evmaddress}
\bibfield{author}{\bibinfo{person}{Vechain}.} \bibinfo{year}{2023}\natexlab{}.
\newblock \bibinfo{title}{Contract address prediction}.
\newblock \bibinfo{howpublished}{\url{https://docs.vechain.org/core-concepts/evm-compatibility/test-coverage/contract-address-prediction}}.
\newblock


\bibitem[Wang et~al\mbox{.}(2018)]%
        {wang2018overview}
\bibfield{author}{\bibinfo{person}{Shuai Wang}, \bibinfo{person}{Yong Yuan}, \bibinfo{person}{Xiao Wang}, \bibinfo{person}{Juanjuan Li}, \bibinfo{person}{Rui Qin}, {and} \bibinfo{person}{Fei-Yue Wang}.} \bibinfo{year}{2018}\natexlab{}.
\newblock \showarticletitle{An Overview of Smart Contract: Architecture, Applications, and Future Trends}. In \bibinfo{booktitle}{\emph{2018 IEEE Intelligent Vehicles Symposium (IV)}}. \bibinfo{pages}{108--113}.
\newblock
\urldef\tempurl%
\url{https://doi.org/10.1109/IVS.2018.8500488}
\showDOI{\tempurl}


\bibitem[Wang et~al\mbox{.}(2024a)]%
        {wang2024efficiently}
\bibfield{author}{\bibinfo{person}{Zexu Wang}, \bibinfo{person}{Jiachi Chen}, \bibinfo{person}{Yanlin Wang}, \bibinfo{person}{Yu Zhang}, \bibinfo{person}{Weizhe Zhang}, {and} \bibinfo{person}{Zibin Zheng}.} \bibinfo{year}{2024}\natexlab{a}.
\newblock \showarticletitle{Efficiently Detecting Reentrancy Vulnerabilities in Complex Smart Contracts}.
\newblock \bibinfo{journal}{\emph{Proc. ACM Softw. Eng.}} \bibinfo{volume}{1}, \bibinfo{number}{FSE}, Article \bibinfo{articleno}{8} (\bibinfo{date}{jul} \bibinfo{year}{2024}), \bibinfo{numpages}{21}~pages.
\newblock
\urldef\tempurl%
\url{https://doi.org/10.1145/3643734}
\showDOI{\tempurl}


\bibitem[Wang et~al\mbox{.}(2024b)]%
        {Unityis}
\bibfield{author}{\bibinfo{person}{Zexu Wang}, \bibinfo{person}{Jiachi Chen}, \bibinfo{person}{Peilin Zheng}, \bibinfo{person}{Yu Zhang}, \bibinfo{person}{Weizhe Zhang}, {and} \bibinfo{person}{Zibin Zheng}.} \bibinfo{year}{2024}\natexlab{b}.
\newblock \showarticletitle{Unity is Strength: Enhancing Precision in Reentrancy Vulnerability Detection of Smart Contract Analysis Tools}.
\newblock \bibinfo{journal}{\emph{IEEE Transactions on Software Engineering}} (\bibinfo{year}{2024}), \bibinfo{pages}{1--12}.
\newblock
\urldef\tempurl%
\url{https://doi.org/10.1109/TSE.2024.3427321}
\showDOI{\tempurl}


\bibitem[{Wikipedia contributors}(2024a)]%
        {bsc}
\bibfield{author}{\bibinfo{person}{{Wikipedia contributors}}.} \bibinfo{year}{2024}\natexlab{a}.
\newblock \bibinfo{title}{Binance --- {Wikipedia}{,} The Free Encyclopedia}.
\newblock \bibinfo{howpublished}{\url{https://en.wikipedia.org/w/index.php?title=Binance&oldid=1214652916}}.
\newblock


\bibitem[{Wikipedia contributors}(2024b)]%
        {ethereum}
\bibfield{author}{\bibinfo{person}{{Wikipedia contributors}}.} \bibinfo{year}{2024}\natexlab{b}.
\newblock \bibinfo{title}{Ethereum --- {Wikipedia}{,} The Free Encyclopedia}.
\newblock \bibinfo{howpublished}{\url{https://en.wikipedia.org/w/index.php?title=Ethereum&oldid=1214478940}}.
\newblock


\bibitem[{Wikipedia contributors}(2024c)]%
        {polygon}
\bibfield{author}{\bibinfo{person}{{Wikipedia contributors}}.} \bibinfo{year}{2024}\natexlab{c}.
\newblock \bibinfo{title}{Polygon (blockchain) --- {Wikipedia}{,} The Free Encyclopedia}.
\newblock \bibinfo{howpublished}{\url{https://en.wikipedia.org/w/index.php?title=Polygon_(blockchain)&oldid=1214411541}}.
\newblock


\bibitem[Wood et~al\mbox{.}(2014)]%
        {wood2014ethereum}
\bibfield{author}{\bibinfo{person}{Gavin Wood} {et~al\mbox{.}}} \bibinfo{year}{2014}\natexlab{}.
\newblock \showarticletitle{Ethereum: A secure decentralised generalised transaction ledger}.
\newblock \bibinfo{journal}{\emph{Ethereum project yellow paper}} \bibinfo{volume}{151}, \bibinfo{number}{2014} (\bibinfo{year}{2014}), \bibinfo{pages}{1--32}.
\newblock


\bibitem[Wood and Wood(2008)]%
        {card_sorting}
\bibfield{author}{\bibinfo{person}{Jed~R Wood} {and} \bibinfo{person}{Larry~E Wood}.} \bibinfo{year}{2008}\natexlab{}.
\newblock \showarticletitle{Card sorting: current practices and beyond}.
\newblock \bibinfo{journal}{\emph{Journal of Usability Studies}} \bibinfo{volume}{4}, \bibinfo{number}{1} (\bibinfo{year}{2008}), \bibinfo{pages}{1--6}.
\newblock


\bibitem[Xue et~al\mbox{.}(2021)]%
        {ye2020clairvoyance}
\bibfield{author}{\bibinfo{person}{Yinxing Xue}, \bibinfo{person}{Mingliang Ma}, \bibinfo{person}{Yun Lin}, \bibinfo{person}{Yulei Sui}, \bibinfo{person}{Jiaming Ye}, {and} \bibinfo{person}{Tianyong Peng}.} \bibinfo{year}{2021}\natexlab{}.
\newblock \showarticletitle{Cross-contract static analysis for detecting practical reentrancy vulnerabilities in smart contracts} \emph{(\bibinfo{series}{ASE '20})}. \bibinfo{publisher}{Association for Computing Machinery}, \bibinfo{address}{New York, NY, USA}, \bibinfo{pages}{1029–1040}.
\newblock
\showISBNx{9781450367684}
\urldef\tempurl%
\url{https://doi.org/10.1145/3324884.3416553}
\showDOI{\tempurl}


\bibitem[Yang et~al\mbox{.}(2023)]%
        {yang2023definition}
\bibfield{author}{\bibinfo{person}{Shuo Yang}, \bibinfo{person}{Jiachi Chen}, {and} \bibinfo{person}{Zibin Zheng}.} \bibinfo{year}{2023}\natexlab{}.
\newblock \showarticletitle{Definition and Detection of Defects in NFT Smart Contracts} \emph{(\bibinfo{series}{ISSTA 2023})}. \bibinfo{publisher}{Association for Computing Machinery}, \bibinfo{address}{New York, NY, USA}, \bibinfo{pages}{373–384}.
\newblock
\showISBNx{9798400702211}
\urldef\tempurl%
\url{https://doi.org/10.1145/3597926.3598063}
\showDOI{\tempurl}


\bibitem[Zhang et~al\mbox{.}(2024)]%
        {jiashuo}
\bibfield{author}{\bibinfo{person}{Jiashuo Zhang}, \bibinfo{person}{Jiachi Chen}, \bibinfo{person}{Zhiyuan Wan}, \bibinfo{person}{Ting Chen}, \bibinfo{person}{Jianbo Gao}, {and} \bibinfo{person}{Zhong Chen}.} \bibinfo{year}{2024}\natexlab{}.
\newblock \showarticletitle{When Contracts Meets Crypto: Exploring Developers' Struggles with Ethereum Cryptographic APIs}. In \bibinfo{booktitle}{\emph{Proceedings of the IEEE/ACM 46th International Conference on Software Engineering}} (Lisbon, Portugal) \emph{(\bibinfo{series}{ICSE '24})}. \bibinfo{publisher}{Association for Computing Machinery}, \bibinfo{address}{New York, NY, USA}, Article \bibinfo{articleno}{164}, \bibinfo{numpages}{13}~pages.
\newblock
\showISBNx{9798400702174}
\urldef\tempurl%
\url{https://doi.org/10.1145/3597503.3639131}
\showDOI{\tempurl}


\bibitem[Zou et~al\mbox{.}(2021)]%
        {zou2019smart}
\bibfield{author}{\bibinfo{person}{Weiqin Zou}, \bibinfo{person}{David Lo}, \bibinfo{person}{Pavneet~Singh Kochhar}, \bibinfo{person}{Xuan-Bach~Dinh Le}, \bibinfo{person}{Xin Xia}, \bibinfo{person}{Yang Feng}, \bibinfo{person}{Zhenyu Chen}, {and} \bibinfo{person}{Baowen Xu}.} \bibinfo{year}{2021}\natexlab{}.
\newblock \showarticletitle{Smart Contract Development: Challenges and Opportunities}.
\newblock \bibinfo{journal}{\emph{IEEE Transactions on Software Engineering}} \bibinfo{volume}{47}, \bibinfo{number}{10} (\bibinfo{year}{2021}), \bibinfo{pages}{2084--2106}.
\newblock
\urldef\tempurl%
\url{https://doi.org/10.1109/TSE.2019.2942301}
\showDOI{\tempurl}


\end{thebibliography}

\end{document}